\documentclass[preprint,onecolumn,nofootinbib]{revtex4}
\pdfoutput=1
\usepackage[colorlinks=true,linkcolor=blue,urlcolor=blue,filecolor=black,citecolor=red,pdfstartview=FitV,pdftitle={},pdfsubject={},pdfkeywords={},pdfpagemode=None,bookmarksopen=true]{hyperref}
\usepackage{graphicx}
\usepackage{amsmath}
\usepackage{amsfonts}
\usepackage{amssymb,ulem}
\usepackage{color}%
\usepackage{dcolumn}
\usepackage{MnSymbol,wasysym}
\usepackage{braket}

\usepackage{eurosym}
\usepackage{calrsfs}
\usepackage[usenames,dvipsnames,svgnames]{xcolor}

\newcommand{\RNum}[1]{\uppercase\expandafter{\romannumeral #1\relax}}

\begin{document}
\baselineskip=0.5 cm
\title{Holographic subregion complexity under thermal quench in Einstein-Maxwell-Axions theory with momentum relaxation}

\author{Yu-Ting Zhou$ ^{1,2}$}
\email{constaantine@163.com}
\author{Xiao-Mei Kuang$ ^{2,4}$}
\email{xmeikuang@yzu.edu.cn}
\author{Yong-Zhuang Li$ ^{3}$}
\email{liyongzhuang@just.edu.cn}
\author{Jian-Pin Wu$ ^{2,4}$}
\email{jianpinwu@yzu.edu.cn}
\affiliation{$^1$ College of Mathematics and Science, Yangzhou University, Yangzhou 225009, China}
\affiliation{$^2$ Center for Gravitation and Cosmology, College of Physical Science and Technology,
Yangzhou University, Yangzhou 225009, China}
\affiliation{$^3$School of Science, Jiangsu University of Science and Technology, Zhenjiang 212003, China}
\affiliation{$^4$ School of Aeronautics and Astronautics, Shanghai Jiao Tong University, Shanghai 200240, China}
\date{\today }

\begin{abstract}
\vspace*{0.6cm}
\baselineskip=0.5 cm
We investigate the evolution of holographic entanglement entropy (HEE) and holographic complexity (HC) under a thermal quench in Einstein-Maxwell-Axion theory (EMA), which is dual to a field theory with momentum relaxation on the boundary. A strip-shaped boundary geometry is utilized to calculate HEE and HC via `entropy=surface' and `complexity=volume' conjecture, respectively. By fixing other parameters we claim that either large enough black hole charge or width of the strip will introduce swallow-tail behaviors in HEE and multi-values in HC due to the discontinuity of the minimum Hubeny-Rangamani-Takayanagi (HRT) surface. Meanwhile, we explore the effects of momentum relaxation on the evolution of HEE and HC. The results present that the momentum relaxation will suppress the discontinuity to occur as it increases. For large enough momentum relaxation the continuity of HEE and HC will be recovered.
\end{abstract}

\maketitle

\tableofcontents

\section{Introduction}

Holography \cite{Maldacena:1997re,Gubser:1998bc, Witten:1998qj} has provided a close connection among the quantum information, condensed matter and quantum gravity. These connections  become more and more important in the realm of theoretical physics and quantum information.
From the view of technics, it is also a powerful tool to study many physical quantities in these areas, especially in strongly correlated systems.

Among the physical quantities, entanglement entropy (EE)  and complexity are two  significant concepts in the theoretical physics and quantum information.
Essentially, EE measures the degrees of freedom in a strongly coupled system while the complexity measures the difficulty of turning a quantum state into another state. But it is extremely difficult to evaluate them on the side of the field theory when the degrees of freedom of the system become large. Fortunately,  both of them can be evaluated with the help of holography and their elegant geometric duality from gravity side have been provided.
Specially, in the holographic framework,  it has been proposed that in \cite{Takayanagi:2012kg,Hubeny:2007xt}, the EE for a subregion on the dual boundary is proportional to the minimal Hubeny-Rangamani-Takayanagi (HRT) surface in the bulk geometry.  Later, two different methods were proposed to evaluate the complexity from geometry. One is the `Complexity=Volume' conjecture (CV), which  states that the holographic complexity (HC) is proportional to the volume of
a codimension-one hypersurface with the AdS boundary and the HRT surface \cite{Stanford:2014jda,Susskind:2014jwa}.  While the other is  `Complexity=Action' conjecture (CA) in which one identifies the HC with the gravitational action evaluated on the Wheeler-DeWitt patch in the bulk \cite{Brown:2015bva,Brown:2015lvg}. Some analytical treatment in this setup has been addressed in \cite{Bhattacharya:2019zkb,Braccia:2019xxi}. In this paper, we will follow the CV conjecture and study its evolution under a thermal quench.

In the framework of holography, some of related quantities and their  geometric descriptions have been holographically investigated. For instance, entanglement of purification (EoP), which is an important quantum information quantity for mixed states, has been holographically dual to the minimal entanglement wedge cross section \cite{Tarhel:0202044, Takayanagi:2017knl,Nguyen:2017yqw} and been generalized in \cite{Umemoto:2018jpc,Liu:2019qje,Ghodrati:2019hnn}.  The bit thread formalism for studying EoP was then addressed in \cite{Bao:2019wcf,Harper:2019lff,Ghodrati:2019hnn,Du:2019emy}.  Complexity of purification (CoP), which describes the minimum number of gates needed to purify a mixed state,  was holographically explored in \cite{Ghodrati:2019hnn,Agon:2018zso}. Another interesting quantity is the logarithmic negativity which captures the quantum correlations with the nature of  entanglement \cite{Plenio2005}. It  is also a quantum entanglement measure for mixed quantum states and the  holographic dual has been studied in \cite{Kudler-Flam:2018qjo, Kusuki:2019zsp}.

Additionally, the study of HEE and HC will provide us more indirect but effective information to explore the nature of the spacetime, in particular the physics of the black hole horizon and its thermal and entanglement structures. Specifically, it was shown in \cite{Susskind:2014rva, Susskind:2014moa} that  the EE is not enough to understand the physics of the black hole horizon via studying the information paradox in black holes.  Therefore, the authors proposed the ER=EPR conjecture \cite{Maldacena:2013xja} and argued that the creation of the firewall behind the horizon is essentially a problem of  quantum computational complexity \cite{Stanford:2014jda,Susskind:2014jwa}, where ER and EPR stand for Einstein-Rosen bridges and Einstien-Podolsky-Rosen correlations, respectively.
The evolution of the HEE and HC under a thermal quench has been explored in various dynamical backgrounds such as Einstein theory \cite{Chen:2018mcc,Auzzi:2019mah,Ling:2019ien},  Einstein-Born-Infeld theory \cite{Ling:2018xpc}, massive gravity theory \cite{Zhou:2019jlh}, and has been further generalized to chaotic system \cite{Yang:2019vgl} and dS boundary \cite{Zhang:2019vgl}. The evolution of HEE and HC for quantum quench have also been investigated in \cite{Leichenauer:2015xra,Leichenauer:2016rxw,Moosa:2017yiz,Fan:2018xwf,Chu:2019etd} and references therein.

In this paper, we will investigate the evolution of HEE and HC under a thermal quench in Einstein-Maxwell gravity coupled with two linear spacial-dependent scalar fields in the bulk,
which is called Einstein-Maxwell-Axions (EMA) theory. The analytical black brane solution with dimension $D\geq 4$ was constructed in \cite{Andrade:2013gsa} and it was found that four dimensional black brane geometry is those of a sector of massive gravity\cite{deRham:2010kj}. Moreover, the evolution of HEE and HC for different dimensions of Einstein theory has been studied in \cite{Chen:2018mcc}. So here we shall focus on the simplest four dimensional case in this setup which is dual to three dimensional boundary theory.
It was addressed in \cite{Andrade:2013gsa} that the scalar fields in the bulk source a spatially dependent field theory with momentum relaxation, while the linear coefficient of the scalar fields  describes the strength of the momentum relaxation.
This means that our study will involve in  momentum relaxation, which is more closer to the reality. Moreover, the effect of  momentum relaxation on  the time evolution of the optical conductivity\cite{Bagrov:2017tqn} and the equilibrium chiral magnetic effect \cite{Fernandez-Pendas:2019rkh}  with Vaidya quench in this model has been studied, which shed light on the quark gluon plasma produced in heavy ion collisions as well as the real-world systems.
It is worthwhile to point out that in this model, momentum relaxation in the dual boundary is sourced by the bulk scalar fields in the bulk, but the black hole geometry  is homogeneous as we will present soon. The fully inhomogeneous holographic thermalization process with spacial dependent bulk geometry  has been studied in \cite{Balasubramanian:2013rva}.

Our paper is organized as follows. In section \ref{sec:BG}, we study the generalized Vaidya-AdS black brane in EMA theory. Then, in section \ref{sec:setup}, we present the holographic setup of HEE and HC for a stripe geometry. We show our results and analyze the effect of momentum relaxation on the evolution of HEE and HC in section \ref{sec:results}.  Finally, section \ref{sec:conclu} contributes to our conclusions and discussions. In this paper, we will set the units as $G=\hbar=c=1$.

\section{Vaidya AdS black branes in Einstein-Maxwell-Axions gravity theory}\label{sec:BG}
We consider the AdS black branes in EMA gravity proposed  in \cite{Andrade:2013gsa}. The action of the four dimensional theory  is given by
\begin{equation}
S=\frac{1}{16\pi }\int \! d^4x \sqrt{-g} \left(R+\frac{6}{\ell^2}-\frac{1}{4}F_{\mu\nu}F^{\mu\nu}-\frac{1}{2}\sum_{I=1}^{2}(\partial\psi_I)^2\right)\ .
\label{eq:action}
\end{equation}
By setting the scalar fields to linearly depend on the two dimensional spatial coordinates $x^a$, i.e., $\psi_I=\beta\delta_{Ia}x^a$ where the index $a$ goes $a=1,2$, the action admits the charged black brane solution
 \begin{eqnarray}\label{eq-metric}
&&ds^2=-r^{2}f(r)dt^2+\frac{1}{r^{2}f(r)}dr^2+r^2(dx^2+dy^2),~~~~~A=A_t(r) dt,~~\mathrm{with}\nonumber\\
&&f(r)=\frac{1}{\ell^2}-\frac{\beta^2}{2\, r^{2}}-\frac{m}{r^3}+\frac{q^{2}}{r^{4}}, ~~~A_t=\left(1-\frac{r_h}{r}\right)\frac{2q}{r_h}.
 \end{eqnarray}
Here, the horizon $r_h$ satisfies $f(r_h)=0$; $\ell$ describes the radius of AdS spacetime, and for  simplicity we will set $\ell=1$. $m$ and $q$ are the mass and charge of the black brane,respectively, with the relation given by
\begin{equation}
1-\frac{\beta^{2}}{2r_{h}^{2}}-\frac{m}{r_{h}^{3}}+\frac{q^{2}}{r_{h}^{4}}=0.
\label{massandcharge}
\end{equation}
It is worthwhile to point out that the scalar fields in the bulk source a spatially dependent boundary field theory with momentum relaxation, which is dual to a homogeneous and isotropic black brane \eqref{eq-metric}. The linear coefficient $\beta$ of the scalar fields is usually considered to describe the strength of the momentum relaxation in the dual boundary theory \cite{Andrade:2013gsa}. A general action with axions terms and the holography has been studied in \cite{Alberte:2015isw}. We note  that the extended thermodynamics of the black brane has also been studied in \cite{Fang:2017nse,Cisterna:2018jqg,Hu:2019wwn}. The Hawking temperature and the thermal entropy density of the black brane reads
\begin{eqnarray}\label{adstemperature}
T=\frac{1}{4 \pi}\frac{d\left(r^{2}f(r)\right)}{dr}{\mid_{r_h}},~~~~s=4\pi r_h^2,
\end{eqnarray}
which is treated as the temperature of the dual boundary field.

With properly chosen coordinate transformation, the above black hole brane $\eqref{eq-metric}$ can be represented as in the Eddington-Finkelstein coordinates,
\begin{eqnarray}
&&ds^{2}=\frac{1}{z^2}\left[-f(z)dv^2-2dvdz+dx^2+dy^2\right],\label{metriads1}\\
&&f(z)=1-\frac{1}{2}\beta^{2}z^{2}-mz^{3}+q^{2}z^{4},~~~A_v=2q(1-z)\label{metriads2}
\end{eqnarray}
with
\begin{eqnarray}\label{coortrans}
&&dv=dt-\frac{1}{f(z)}dz \qquad \mathrm{and}\qquad z=\frac{1}{r}.
\end{eqnarray}
We note that the coordinates $v$ and $t$ coincide on the boundary.
Thus, in order to  holographically describe the evolution of HEE and HC, one usually frees the mass and charge  parameter
as smooth  functions of $v$ as \cite{Balasubramanian:2011ur,Galante:2012pv}
\begin{eqnarray}
M(v)&=\frac{m}{2}\left[1+\text{tanh}\left(\frac{v}{v_0}\right)\right],\label{dynamasspara}\\
Q(v)&=\frac{q}{2}\left[1+\text{tanh}\left(\frac{v}{v_0}\right)\right],\label{dynachargepara}
\end{eqnarray}
where $v_0$ represents the finite thickness of the falling charged dust shell. Then the related Vaidya AdS black brane  is
\begin{eqnarray}
&&ds^{2}=\frac{1}{z^2}\left[-f(v, z)dv^2-2dvdz+dx^2+dy^2\right], \label{Vaidyametriads1}\\
&& \mathrm{with}~~ f(v, z)=1-\frac{1}{2}\beta^{2}z^{2}-M(v)z^{3}+Q(v)^{2}z^{4},\label{Vaidyametriads2}\\
\mathrm{and} &&A_v=2Q(v)(1-z).\label{VaidyAv2}
\end{eqnarray}
Now $v$ stands for the ingoing null trajectory, which coincides with the time coordinate $t$ on the conformal boundary. It is easy to check when $v\rightarrow +\infty$, the above formula reduces to the black brane solution \eqref{metriads1}, while in the limit $v\rightarrow -\infty$, it can recover the pure AdS spacetime when $\beta=0$.

Following the strategy of \cite{Caceres:2012em}, we obtain the above solution \eqref{Vaidyametriads1}-\eqref{VaidyAv2} corresponds to the external sources of current and energy-momentum tensor
\begin{eqnarray}
J^z_{(ext)}&=&2\frac{dQ(v)}{dv},\\
T_{vv}^{(ext)}&=&z^2\frac{dM(v)}{dv}-2z^3Q(v)\frac{dQ(v)}{dv},
\end{eqnarray}
respectively.  It is noticed that in order to probe the  time-dependent optical conductivity without  translation invariance, more external sources  were considered to construct the solution in Vaidya setup of the EMA theory\cite{Bagrov:2017tqn}.  However, here in our solution, we consider the simple case that only the mass and charge depend on the time but the momentum relaxation coefficient $\beta$ does not vary with  time. This is reasonable because as addressed in \cite{Fernandez-Pendas:2019rkh} for five dimensional case  that $M$ and $Q$ are integration constants when solving
the differential equations while  $\beta$ is fixed when sourcing the scalars.

\section{Holographic setup of HEE and HC for a stripe}\label{sec:setup}
In this section, we shall address the holographic setup of  HEE and HC under a thermal quench  in the field theory with momentum relaxation, which is dual to the bulk with axions described in last section.
\begin{figure}[ht!]
 \centering
  \includegraphics[width=7cm] {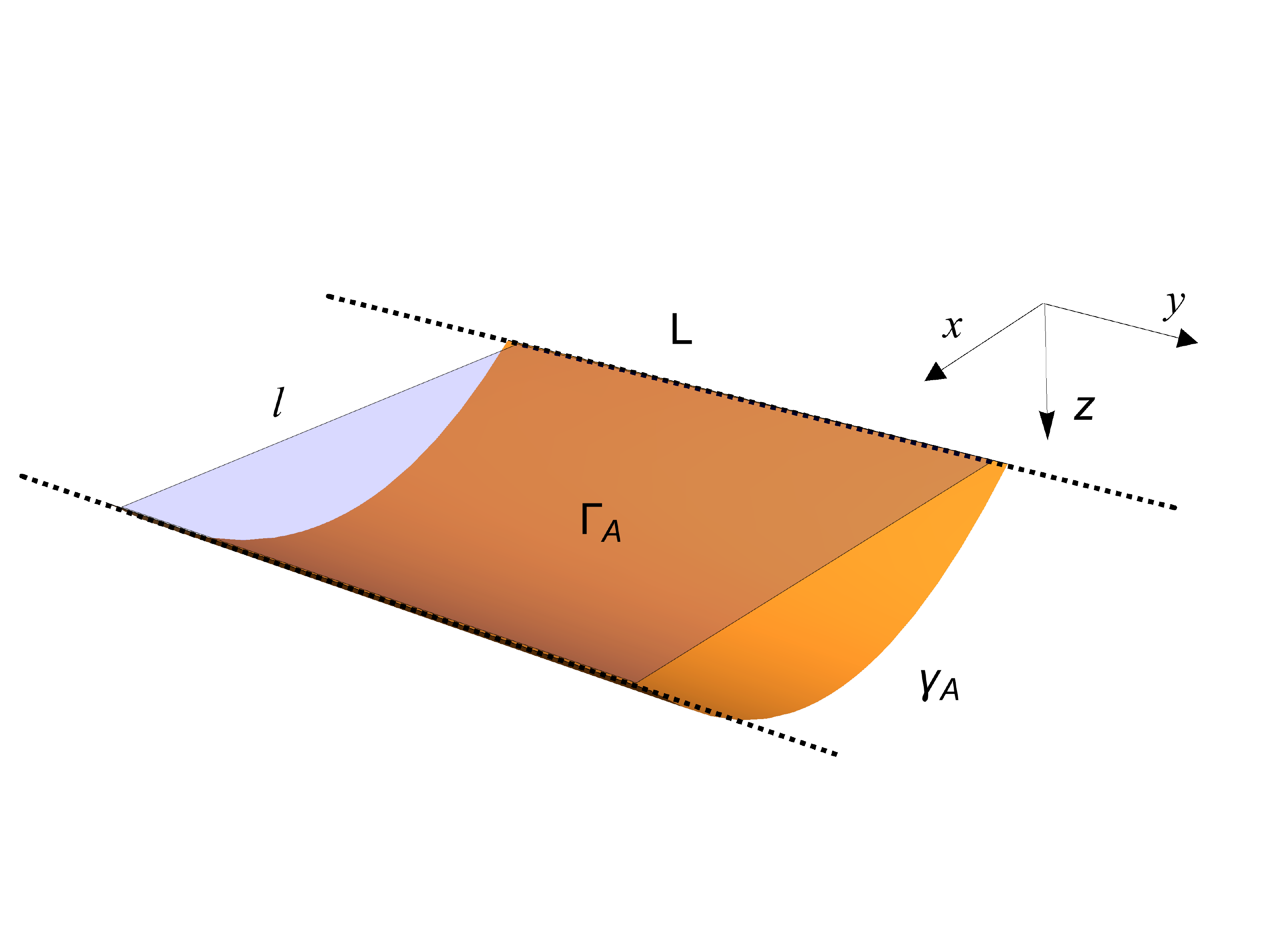}
  \caption{Geometrical description of the subregion $\mathcal{A}$ (the light blue area) with width $l$ and length $L$. The yellow area indicates the HRT bulk surface $\gamma_{\mathcal{A}}$ while
 the  hypersurface $\Gamma_{\mathcal{A}}$ is the bulk area with boundaries $\mathcal{A} $ and $\gamma_{\mathcal{A}}$.
}
 \label{fig:strips}
\end{figure}
We first consider the subregion with a straight strip geometry described by  $\mathcal{A}\equiv\left\{x \in\left(-\frac{l}{2}, \frac{l}{2}\right), y\in\left(-\frac{L}{2}, \frac{L}{2}\right)\right\}$, see Fig.\ref{fig:strips} where we will set the width $l$ to be finite and the length of the region $\mathcal{A}$ to be $L\to \infty$.
As proposed in \cite{Hubeny:2007xt} that in the dynamical spacetime, HEE  for a subregion $\mathcal{A}$ on the boundary is captured by the HRT bulk surface $\gamma_{\mathcal{A}}$, while the corresponding HC
is proportional to the volume of a codimension-one hypersurface $\Gamma_{\mathcal{A}}$ with the boundaries  $\mathcal{A}$ and $\gamma_{\mathcal{A}}$.
Thus, we will follow the steps in \cite{Chen:2018mcc} to analytically deduce the expressions of HEE via the minimal surface, and HC  via CV conjecture of the boundary theory. It is noticed that other probes, such as Wilson Loop and HEE in static case \cite{Mozaffara:2016iwm} and  during thermal quench \cite{Li:2019} in this theory with momentum relaxation have been investigated.

Due to the symmetry of the system, one can parameterize the corresponding extremal surface $\gamma_{\mathcal{A}}$ in the bulk  as
\begin{eqnarray}
v=v(x), \quad z=z(x), \quad z(\pm l / 2)=\epsilon, \quad v(\pm l / 2)=t-\epsilon,
\end{eqnarray}
where $\epsilon$ is the UV cut-off. Then the induced metric on the surface is
\begin{eqnarray}
d s^{2}=\frac{1}{z^{2}}\left[-f(v, z) v^{\prime 2}-2 z^{\prime} v^{\prime}+1\right] d x^{2}+\frac{1}{z^{2}}dy^2,
\end{eqnarray}
where the prime means taking the derivative to $x$. Then the area of the extremal surface is calculated as
\begin{eqnarray}
\operatorname{Area}\left(\gamma_{\mathcal{A}}\right)= L\int_{-l / 2}^{l / 2} \frac{\sqrt{1-f(v, z) v^{\prime 2}-2 z^{\prime} v^{\prime}}}{z^2}d x.
\end{eqnarray}
To obtain the HEE, one has to minimize the above area. The trick is to treat the above  function as an action with $x$ instead of `time', and so the related Lagrangian and Hamiltonian density are
\begin{eqnarray}
\label{eq-lagrangian}\mathcal{L}_{S}&=&\frac{\sqrt{1-f(v, z) v^{\prime 2}-2 z^{\prime} v^{\prime}}}{z^2},\\
\label{eq-Hamitonion}\mathcal{H}_{S}&=&\frac{1}{z^2\sqrt{1-f(v, z) v^{\prime 2}-2 z^{\prime} v^{\prime}}}.
 \end{eqnarray}
It is obvious that the Hamiltonian does not explicitly depend on the variable $x$, so it is conserved. Beside, the symmetry of the surface implies a turning point $(z_{*}, v_{*})$ at $x =0$ on the extremal surface  $\gamma_{\mathcal{A}}$(see Fig.\ref{fig:strips} for a fixed $v$).
Then we can set
\begin{eqnarray}\label{eq-boundarycondition}
 v^{\prime}(0)=z^{\prime}(0)=0, \quad z(0)=z_{*}, \quad v(0)=v_{*}.
\end{eqnarray}
Subsequently, the conserved Hamitonian give us a condition
 \begin{eqnarray}
1-f(v, z) v^{\prime 2}-2 z^{\prime} v^{\prime}=\frac{z_{*}^{4}}{z^{4}}.
 \end{eqnarray}
One can take derivative of the Lagrangian \eqref{eq-lagrangian} with respect to $x$. Combining the derivative equation and the equations of motion for $z(x)$ and for  $v(x)$, respectively, we get a group of partial differential equations
  \begin{eqnarray}
\label{eq-PDE01}0&=&-4+2 z v^{\prime \prime}+v^{\prime}\left[4 f(v, z) v^{\prime}+8 z^{\prime}-z v^{\prime} \partial_{z} f(v, z)\right],\\
\label{eq-PDE02}0&=& 4 f(v, z)^{2} v^{\prime 2}+f(v, z)\left[-4+8 v^{\prime} z^{\prime}-z v^{2} \partial_{z} f(v, z)\right] \nonumber\\ &&~-z\left[2 z^{\prime \prime}+v^{\prime}\left(2 z^{\prime} \partial_{z} f(v, z)+v^{\prime} \partial_{v} f(v, z)\right)\right].
   \end{eqnarray}

By solving the above equations using the boundary conditions \eqref{eq-boundarycondition}, one can extract the solutions of $v=\tilde{v}(x), z=\tilde{z}(x)$ for the extremal surface $\gamma_{\mathcal{A}}$. Then the area of the extremal surface $\gamma_{\mathcal{A}}$ is simplified as
\begin{eqnarray}\label{eq-S0}
\operatorname{Area}\left(\gamma_{\mathcal{A}}\right)=2 L \int_{0}^{l / 2} \frac{z_{*}^2}{\tilde{z}(x)^{4}} d x,
\end{eqnarray}
which gives the HEE of the subregion on the boundary. It is noted that the surface does not live on a constant time slice for the general $f(v,z)$, and both $z_{*}$ and $\tilde{z}(x)$ are time dependent.

Using the same profile as HEE, we then further derive the general expression of HC by evaluating the dual volume in the bulk of the background \eqref{metriads1}, i.e., we should evaluate the volume with the codimension-one extremal surface $\Gamma_{\mathcal{A}}$ which is bounded by the surface $ \gamma_{\mathcal{A}}$.  As discussed in \cite{Chen:2018mcc}, depending on the parametrization, there are two schemes to describe $\Gamma_{\mathcal{A}}$ which is enclosed by  $v=\tilde{v}(x), z=\tilde{z}(x) $.

One scheme is to parameterize $\Gamma_{\mathcal{A}}$ via $ z=z(v)$.
Thus, the induced metric on $\Gamma_{\mathcal{A}}$ is
 \begin{eqnarray}
 d s^{2}=\frac{1}{z^{2}}\left[-\left(f(v, z)+2 \frac{\partial z}{\partial v}\right) d v^{2}+d x^{2}+d y^{2}\right].
 \end{eqnarray}
Consequently,  the volume can be evaluated as
 \begin{eqnarray}\label{eq-V00}
 V(\Gamma_{\mathcal{A}})=2 L \int_{v_{*}}^{\tilde{v}(l / 2)} d v \int_{0}^{\tilde{x}(v)} \frac{d x}{z^3}\left[-f(v, z)-2 \frac{\partial z}{\partial v}\right]^{1 / 2},
 \end{eqnarray}
where $\tilde{x}(v)$ is one coordinate in the codimension-two extremal surface $\gamma_{\mathcal{A}}$.
We treat the above integral function as a Lagrangian,  and then the corresponding  equation of motion is
 \begin{eqnarray}\label{eq-eomForV}
 \begin{aligned} 0=&6f(v, z)^{2}+12z^{\prime}(v)^{2}-3 z(v) z^{\prime}(v) \partial_{z} f(v, z)+f(v, z)\left(18 z^{\prime}(v)-z(v) \partial_{z} f(v, z)\right)\\ &-z(v)\left(2 z^{\prime \prime}(v)+\partial_{v} f(v, z)\right) .
 \end{aligned}
  \end{eqnarray}
There are two possible ways to get the solution to the above equation. One is to solve it with the use of the boundary conditions  determined by the codimension-two surface $\gamma_{\mathcal{A}}=(\tilde{v}(x), \tilde{z}(x))$ and $\mathcal{A}$. The other is to figure out the solution  by seeking $\tilde{z}(\tilde{v})$ on the boundary $\gamma_{\mathcal{A}}$. Subsequently, the volume \eqref{eq-V00} can be further written as
\begin{eqnarray}\label{eq-V0}
V(\Gamma_{\mathcal{A}})=2L  \int_{v_{*}}^{\tilde{v}(l / 2)} \frac{d v}{z(v)^{3}}\left[-f(v, z(v))-2 \frac{\partial z}{\partial v}\right]^{1 / 2}  \tilde{x}(v).
\end{eqnarray}

The other scheme is to parametrize $\Gamma_{\mathcal{A}}$ via $v=v(z)$, in which the induced metric on the codimension-one surface is
 \begin{eqnarray}
 d s^{2}=\frac{1}{z^{2}}\left[-\left(f(v, z)\frac{\partial v}{\partial z}+2 \right) \frac{\partial v}{\partial z} d z^{2}+d x^{2}+d y^{2}\right].
 \end{eqnarray}
Subsequently, following the same step in the former scheme, the on-shell volume is reduced as
\begin{eqnarray}\label{eq-V00}
V(\Gamma_{\mathcal{A}})=2L  \int_{0}^{z_{\star}} \frac{d z}{z^{3}}\left[-f(v(z), z)\left(\frac{\partial v}{\partial z}\right)^2-2 \frac{\partial v}{\partial z}\right]^{1 / 2}  \tilde{x}(z).
\end{eqnarray}

It was pointed out in \cite{Chen:2018mcc} that there are cases that $\tilde{v}(z)$ and $\tilde{x}(z)$ are multi-valued functions of $z$ (this can also be seen in our late study), so the integral \eqref{eq-V00} is not always well defined in the whole process of  evolution. While $\tilde{z}(\tilde{v})$ is a single valued function of $\tilde{v}$ all the time, so the integral \eqref{eq-V0} is well defined during the evolution. Moveover, the expression \eqref{eq-V00} is more intuitive than  \eqref{eq-V0} in the static background, and their outcome is the same when $\tilde{v}(z)$ is singly valued function of $z$. Consequently, we will employ the integral \eqref{eq-V0} to compute the HC in the next discussion.

So far, we have obtained the explicit expressions of HEE  \eqref{eq-S0} and HC \eqref{eq-V0} of the boundary theory dual to the background \eqref{eq-metric} under a thermal quench.  In the following section we will show our numerical results presenting the evolving behaviors of HEE and HC.

\section{Numerical results}\label{sec:results}
To numerically study the evolution of HEE and HC, we have to solve the equations of motion \eqref{eq-PDE01} and \eqref{eq-PDE02}  with the  boundary conditions,
\begin{eqnarray}\label{eq-bdys}
v^{\prime}(0)=z^{\prime}(0)=0\,,\,\, z(0)=z_{*}\,,\,\, v(0)=v_{*}\,,\,\,  z(l / 2)=\epsilon\,,\,\,  v(l / 2)=t-\epsilon\,.
\end{eqnarray}

Conventionally, we are only interested in the finite physical quantities while both HEE and HC evaluated by above conditions are divergent if $\epsilon\rightarrow 0$.
Therefore, we could define the following relative renormalized finite terms
\begin{eqnarray}
&&
S=\frac{\operatorname{Area}(\gamma_{\mathcal{A}})-\operatorname{Area}_{\operatorname{AdS}}(\gamma_{\mathcal{A}})}{2L}
\,,\label{subtractHEE}
\\
&&
C=\frac{\operatorname{V}(\Gamma_{\mathcal{A}})-\operatorname{V}_{\operatorname{AdS}}(\Gamma_{\mathcal{A}})}{2L}\,,
\label{subtractHC}
\end{eqnarray}
where $\operatorname{Area}(\gamma_{\mathcal{A}})$ and $\operatorname{V}(\Gamma_{\mathcal{A}})$ are defined in equation \eqref{eq-S0} and \eqref{eq-V0}, respectively. The quantities with subscript $AdS$ correspond to the vacuum part  dual to the AdS geometry with vanishing mass and charge, which we have to find the minimal surface $\gamma_{\mathcal{A}}$ and the related volume of $\Gamma_{\mathcal{A}}$, separately. Usually they can be integrated out as $\mathrm{Area}_{AdS}(\gamma_{\mathcal{A}})=\frac{2L}{\epsilon}+\frac{L}{2z_{*}}\frac{\sqrt{\pi}\Gamma(-\frac{1}{4})}
{\Gamma(\frac{1}{4})}$ and $\mathrm{V}_{AdS}(\Gamma_{\mathcal{A}})=\frac{Ll}{2\epsilon^2}+\frac{\sqrt{\pi}L}{z_{*}}\left(\frac{2\Gamma(\frac{3}{4})}
{\Gamma(\frac{1}{4})}-\frac{3\Gamma(\frac{1}{4})}
{4\Gamma(\frac{3}{4})}\right)$, respectively.

Next, we  present the numerical results for the evolutions of these two quantities with quench. For the sake of the numerical precision, we will set the UV cutoff $\epsilon=0.05$, the thickness of the shell $v_0=0.01$ and the mass parameter $m=1$ in the calculation.

\subsection{Evolution of HEE and HC in neutral case}
To see the role of  momentum dissipation plays in the evolution of HEE and HC, we first consider the neutral case, i.e., $q=0$
so we can focus on the effect of $\beta$. In addition, we fix the strip width $l=2$ as the first step.

\begin{figure}[ht!]
 \centering
  \includegraphics[width=5cm]{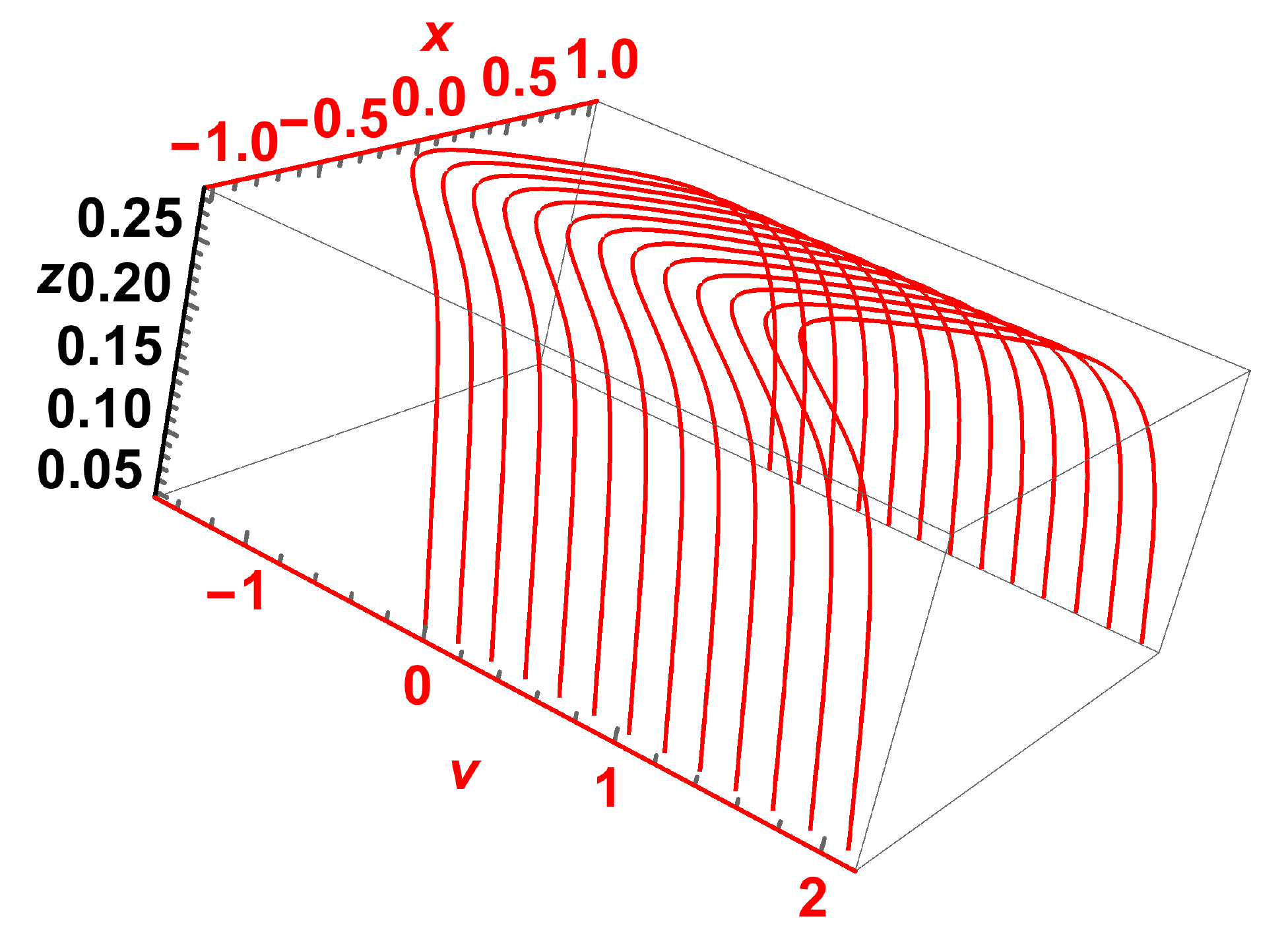}\ \hspace{0.1cm}
  \includegraphics[width=5cm]{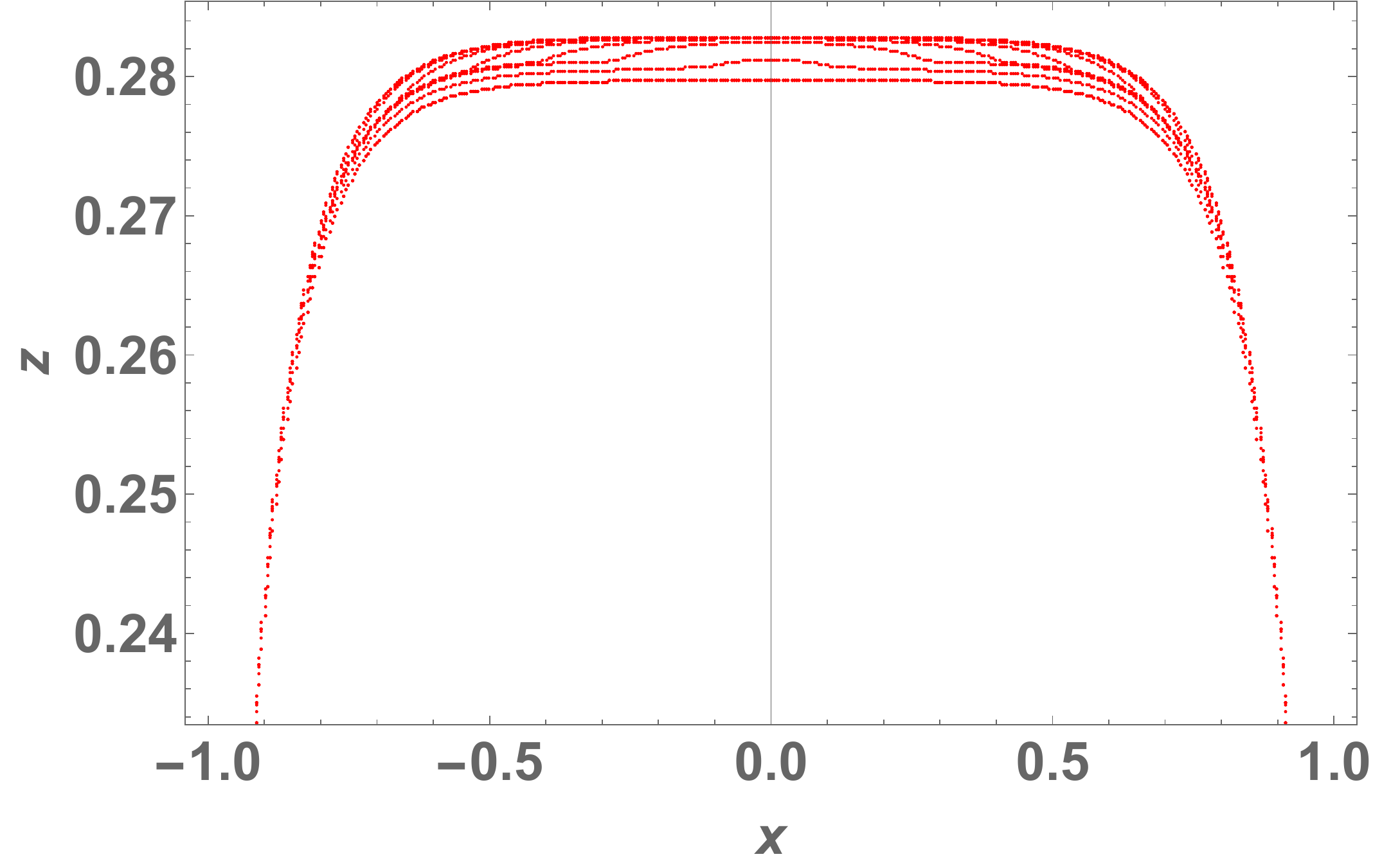}\ \hspace{0.1cm}
  \includegraphics[width=5cm]{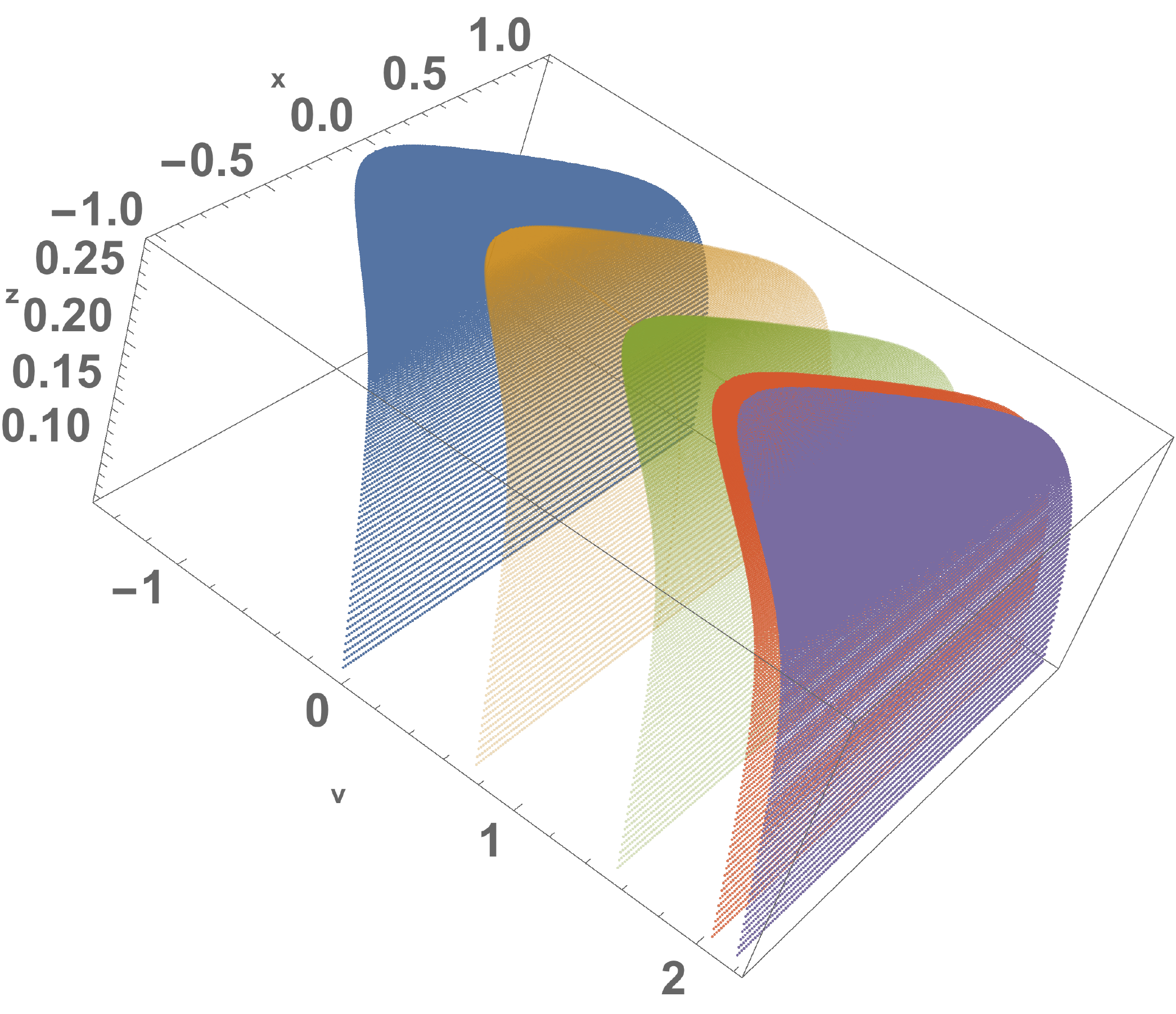}\ \hspace{0.1cm}
        \caption{Left: The evolution of HRT surface in $(x,v,z)$  space.
	 Middle: The projection of HRT surface in $(x, z)$ planer . Right: the volumn of $\Gamma_{\mathcal{A}}$.
	  Here we have set $\beta =5$ and $l=2$.}
 \label{fig:HRTzvxl2}
\end{figure}

Before presenting the main results of HEE and HC,  we first show  the evolution of the HRT surface $\gamma_{\mathcal{A}}$ for $\beta=5$ to give an intuitive understanding of the evolution. The left panel in Fig.\ref{fig:HRTzvxl2} exhibits the evolution in $(x,v,z)$ space, in which $\gamma_{\mathcal{A}}$ evolves from left to right. In the middle panel of Fig.\ref{fig:HRTzvxl2}, the corresponding projection in $(x,z)$ plan is shown, in which the evolution is from top to bottom.
It is obvious that $\gamma_{\mathcal{A}}$ evolves smoothly from the initial state to the final state. After obtaining the HRT surface $\gamma_{\mathcal{A}}$, we can then work out the codimension-one surface $\Gamma_{\mathcal{A}}$, which characterizes the subregion complexity bounded by the HRT surface $\gamma_{\mathcal{A}}$ as well as subregion $\mathcal{A}$, which is exhibited in the right of  Fig.\ref{fig:HRTzvxl2}. The process is similar with the case in the Einstein gravity in \cite{Chen:2018mcc}.
\begin{figure}[ht!]
 \centering
  \includegraphics[width=5cm]{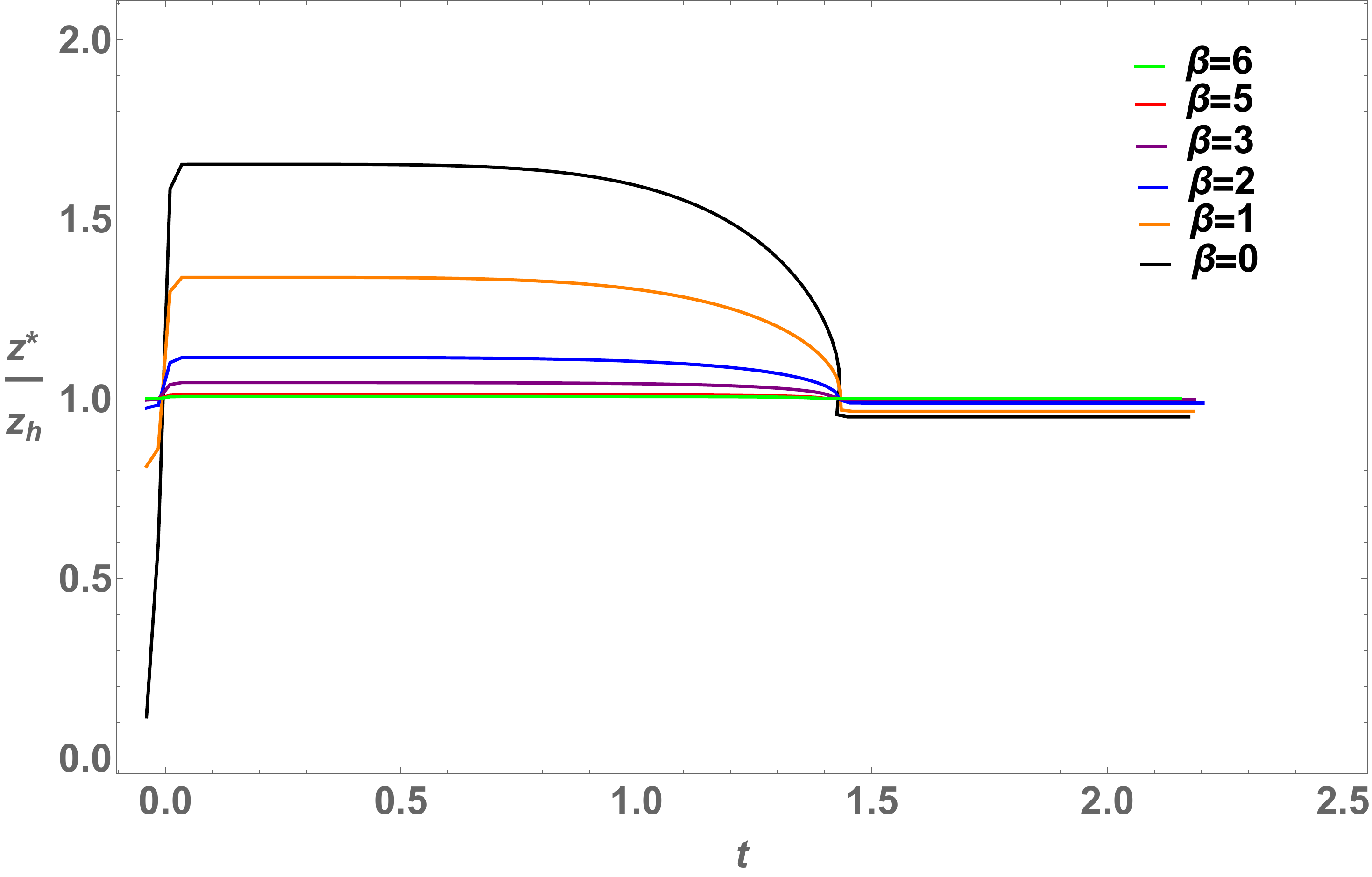}\ \hspace{0.1cm}
  \includegraphics[width=5cm]{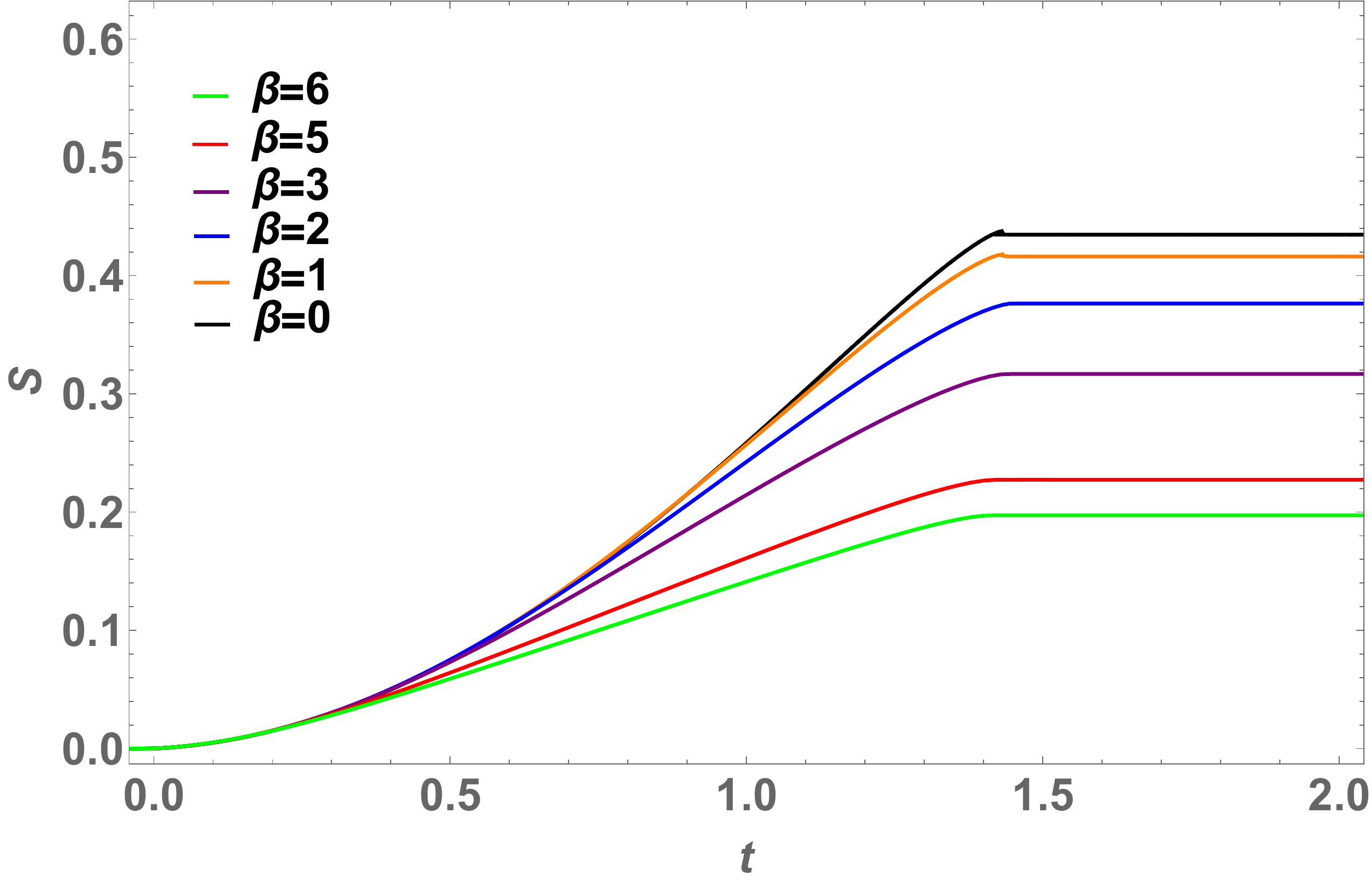}\ \hspace{0.1cm}
  \includegraphics[width=5cm]{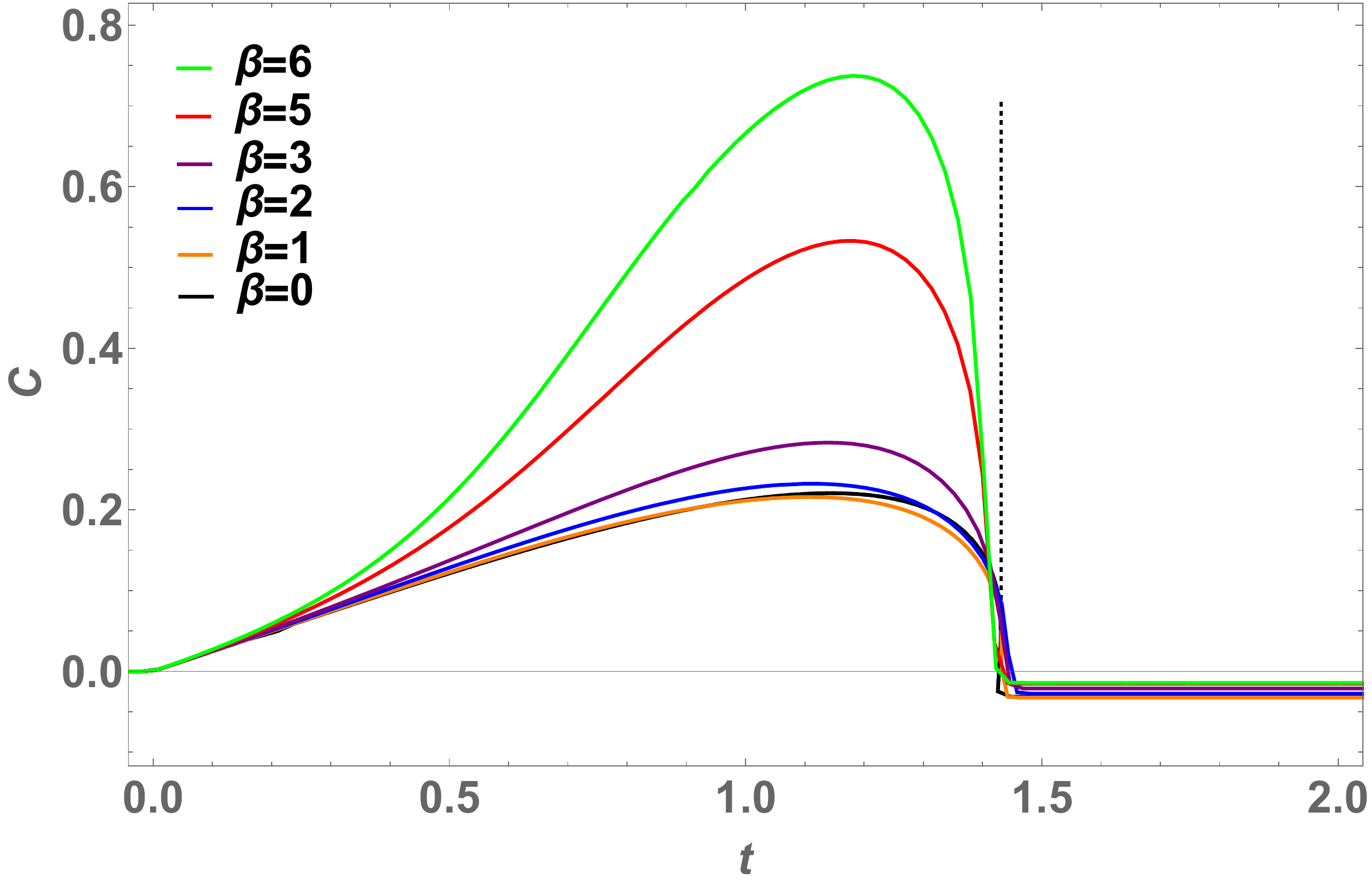}
	  \caption{Left: The evolution of $z_{*}/z_h$ for different $\beta $.
	  Middle: The evolution of HEE  the for different $\beta $. Right: The relation between the HC  and $\beta $.
	  Here we have set $l=2$.}
 \label{fig:q0l2zsc}
\end{figure}

\begin{figure}[ht!]
 \centering
  \includegraphics[width=5cm]{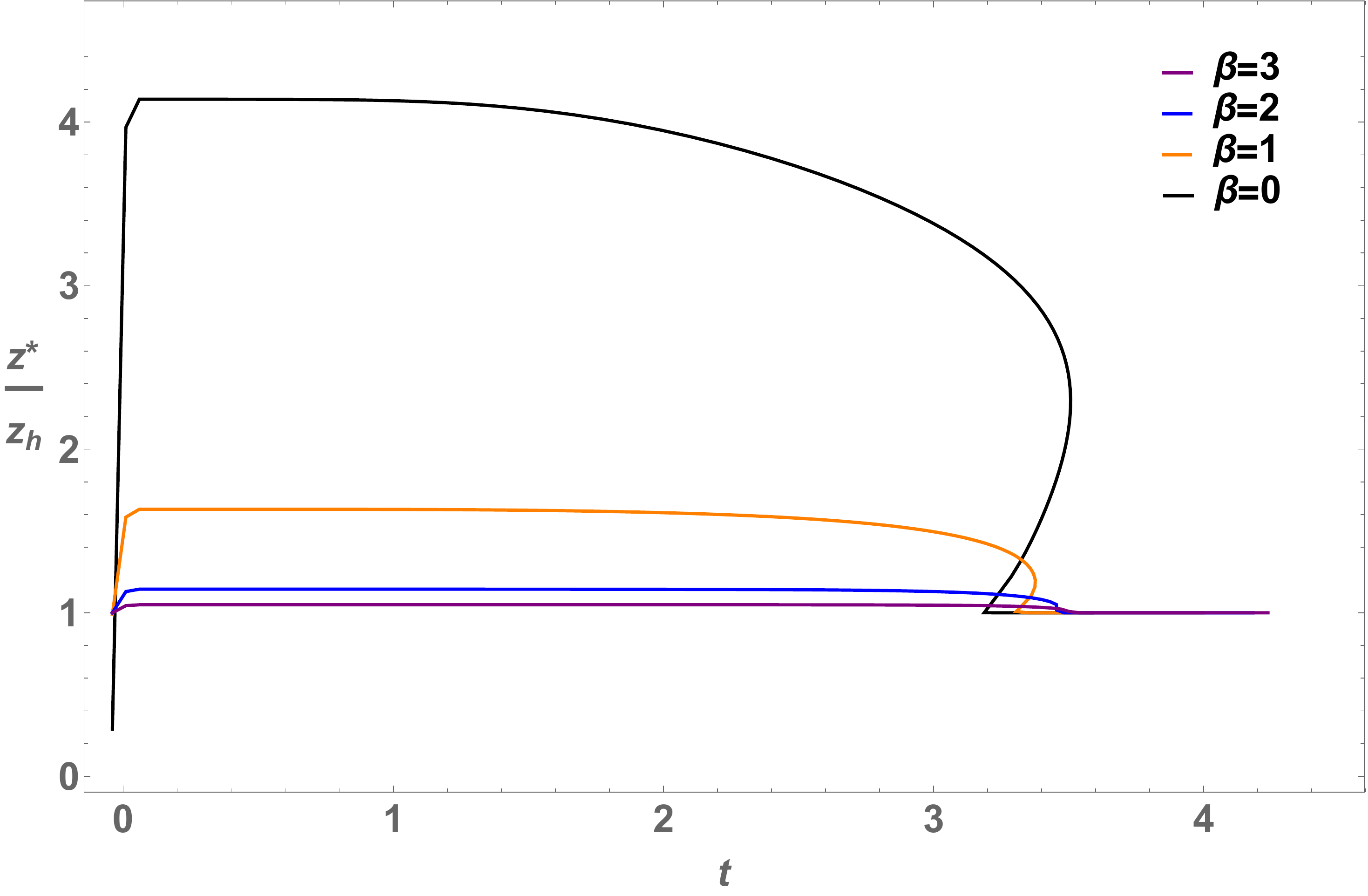}\ \hspace{0.1cm}
  \includegraphics[width=5cm]{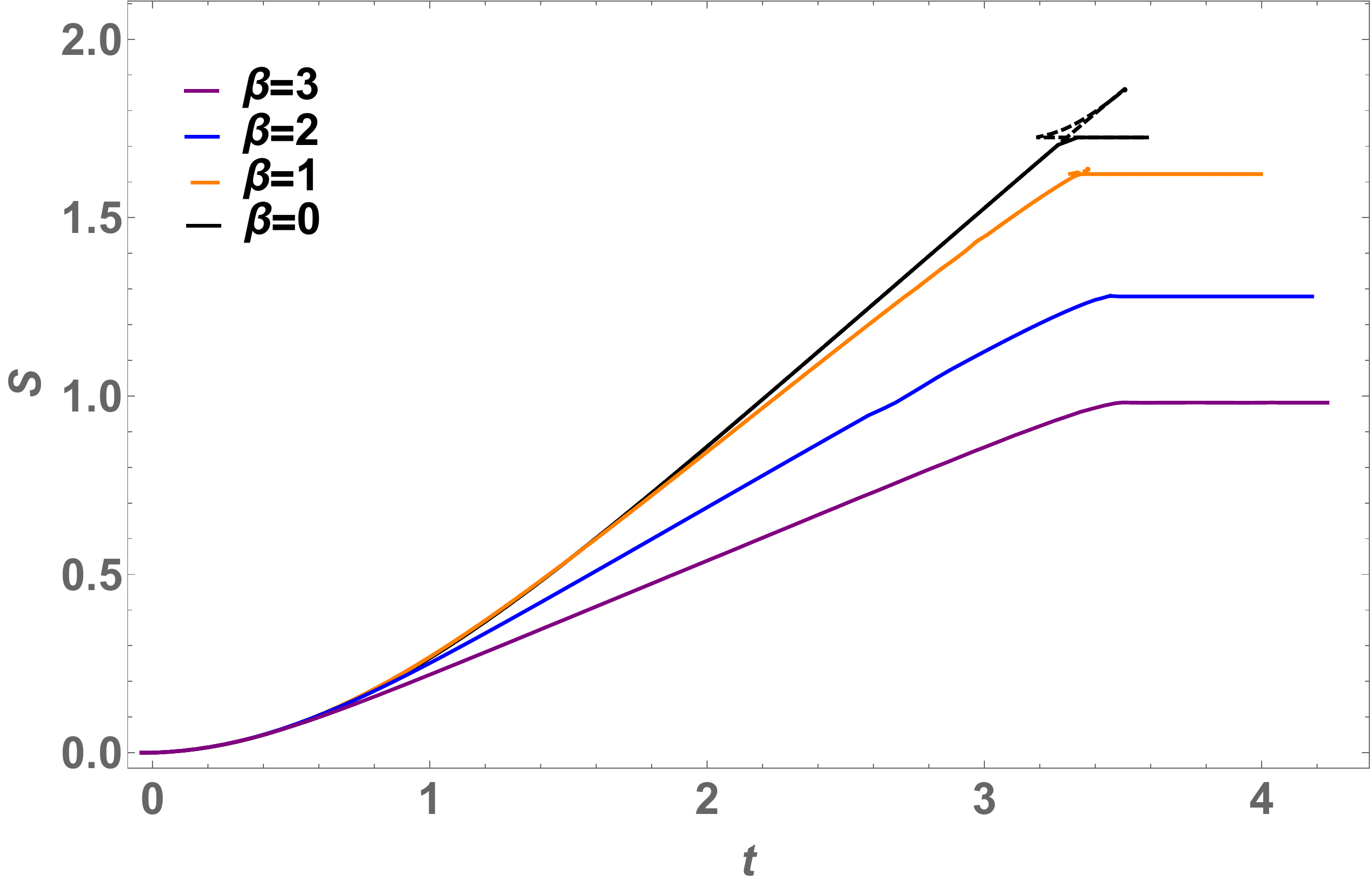}\ \hspace{0.1cm}
  \includegraphics[width=5cm]{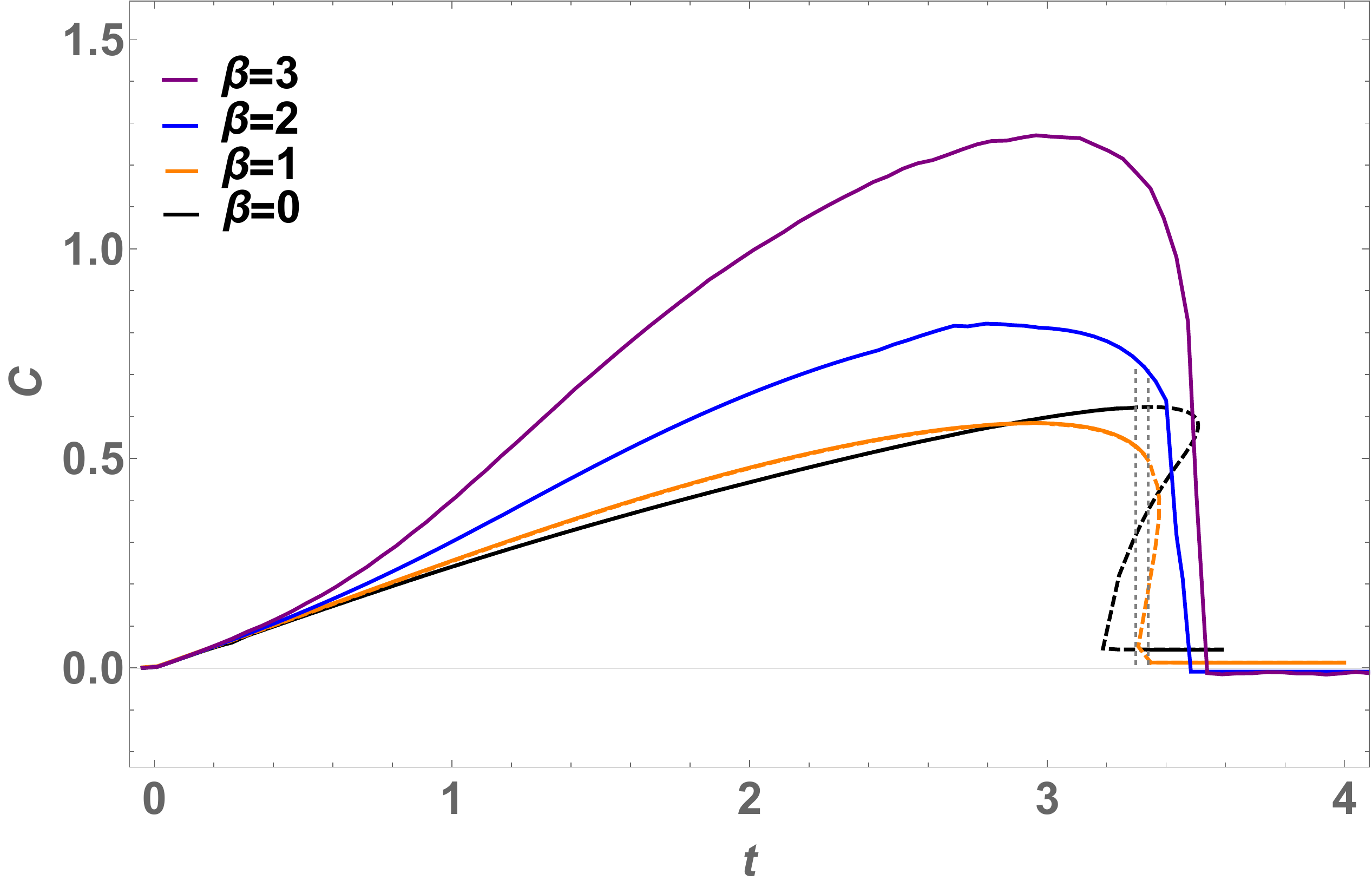}
	  \caption{Left: The evolution of $z_{*}/z_h$ for different $\beta $.
	  Middle: The evolution of HEE for different $\beta $. Right: The relation between the HC  and $\beta $.
	  Here we have set $l=5$.}
 \label{fig:q0l5zsc}
\end{figure}

The evolution of HEE and HC affected by $\beta$ is shown in Fig.\ref{fig:q0l2zsc}. Several properties can be read off from the figures. First, the left panel shows the evolution of the turning point $z_{*}$ over the horizon radius $z_h$. We see that at the beginning of evolution, since $z_h$ decreases sharply as the horizon radius increases suddenly under the quench function \eqref{dynamasspara}, the ratio $z_{*}/z_h$ increases sharply to be larger than $1$, meaning the HRT surface is outside the horizon. As time passes, the ratio drops and approaches to $1$ which implies that the HRT surface is always inside the horizon as the system becomes saturate. The corresponding stable HEE in the middle panel for larger $\beta$ is smaller. In the contrary, the right panel shows that $\beta$ slightly affect the stable value of HC while its peak is explicitly higher.
 We then study the effect of $\beta$ on HEE and HC with wide subregion, i.e., $l=5$. The results are plotted in Fig.\ref{fig:q0l5zsc}.

\begin{figure}[ht!]
 \centering
  \includegraphics[width=7cm]{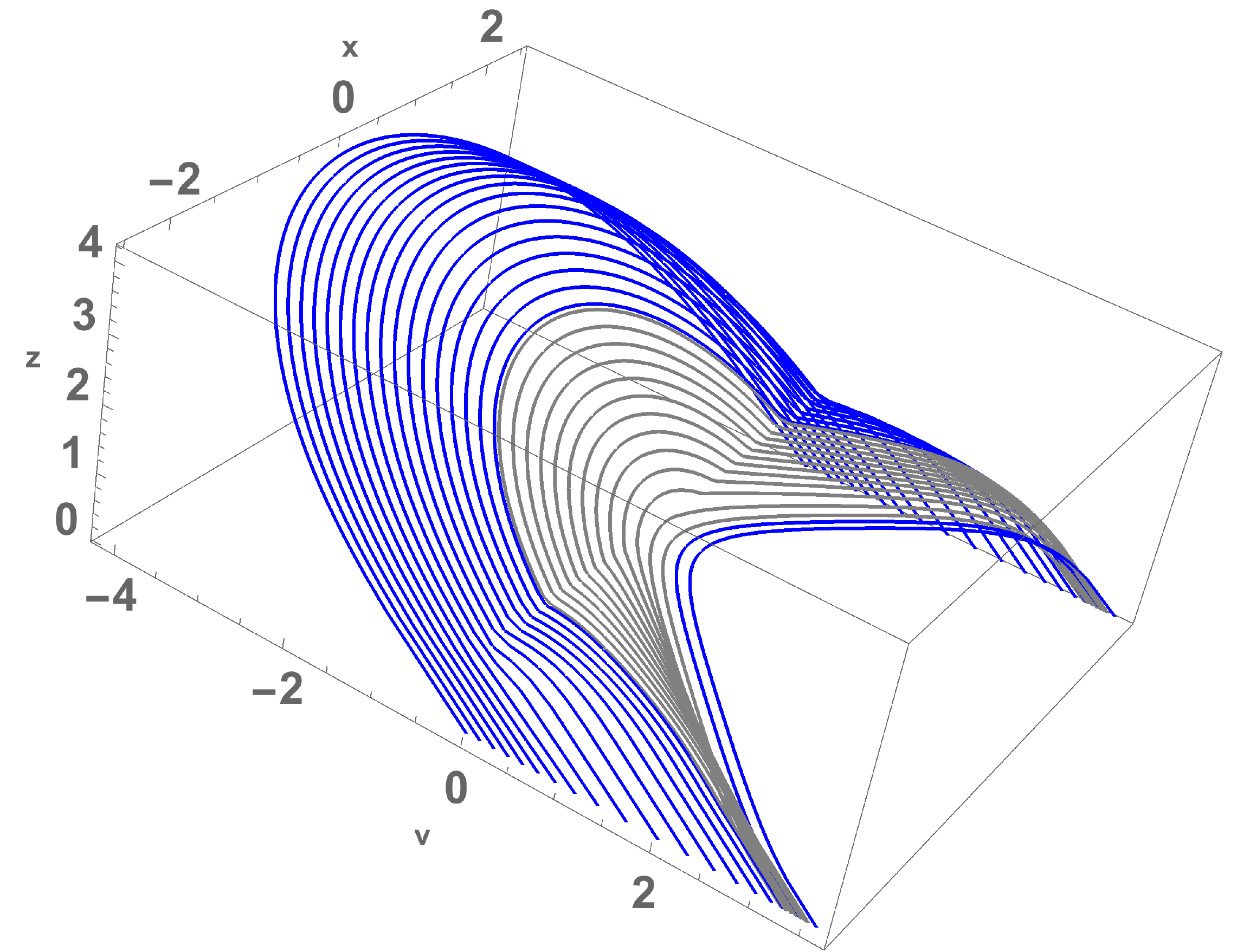}\ \hspace{1cm}
  \includegraphics[width=6cm]{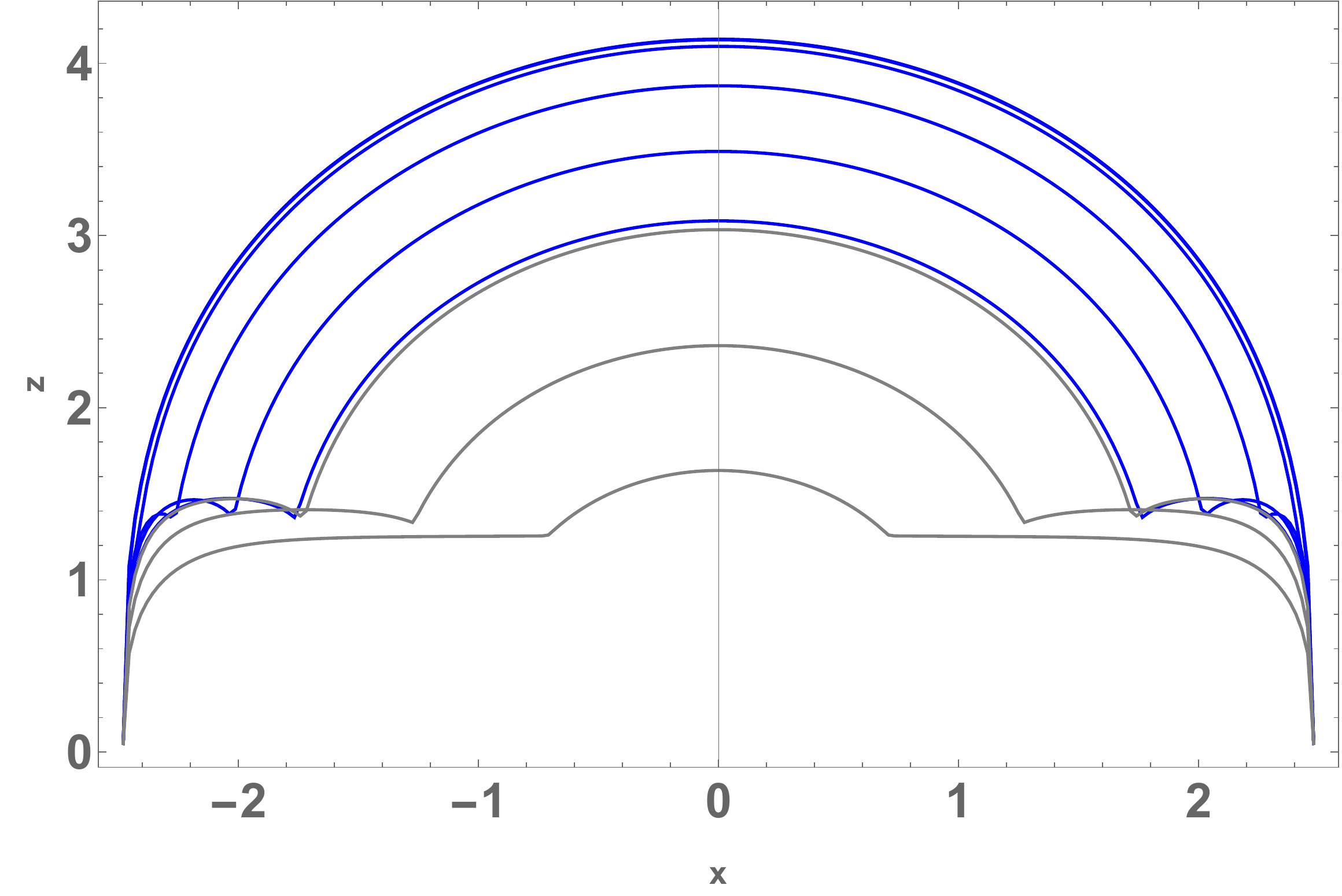}\ \hspace{1cm}
        \caption{Left: The evolution of HRT surface in $(x,v,z)$ space.
	  Right: The projection of HRT surface in $(x, z)$ planer.
	  Here we have set $\beta =0$ and $l=5$.}
 \label{fig:HRTzvxl6}
\end{figure}

For $\beta=0$, we reproduced the result of four dimensional Schwashiz black hole \cite{Chen:2018mcc}, which is denoted by the black lines in each plot. For different $\beta$, HEE has  swallow-tail behaviors  and HC has the multi-valued regions, but only solid lines describe the physical procedure. This phenomena can be explained by the evolution of $\gamma_{\mathcal{A}}$ shown in Fig.\ref{fig:HRTzvxl6}, which is no longer a continuous function and different from the case of the small width strips.
The discontinuous evolution, i.e., the dropping behavior is related with the jump in the minimal area surface from phenomena perspective.

\begin{figure}[ht!]
 \centering
  \includegraphics[width=7cm]{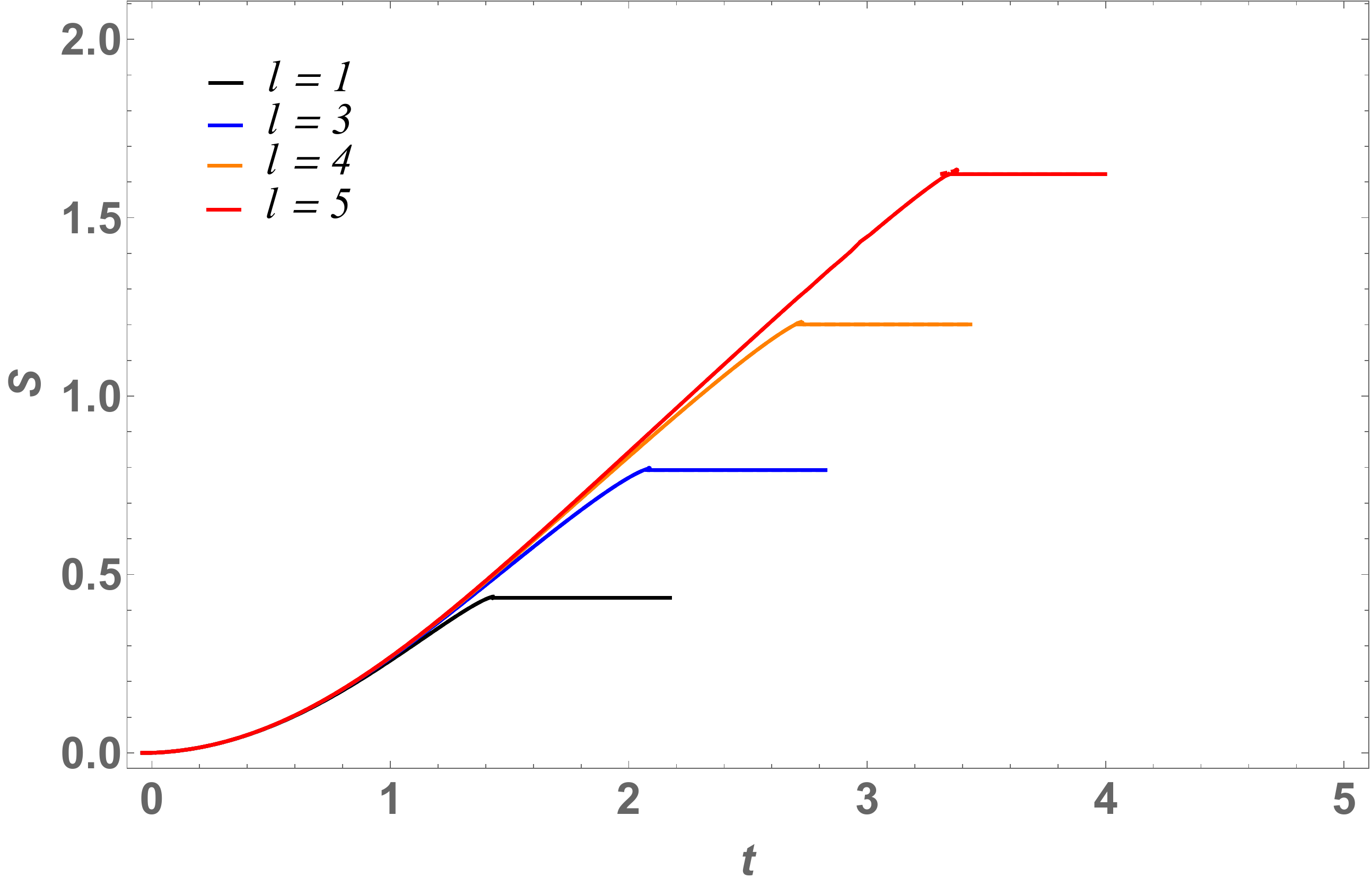}\ \hspace{1cm}
  \includegraphics[width=7cm]{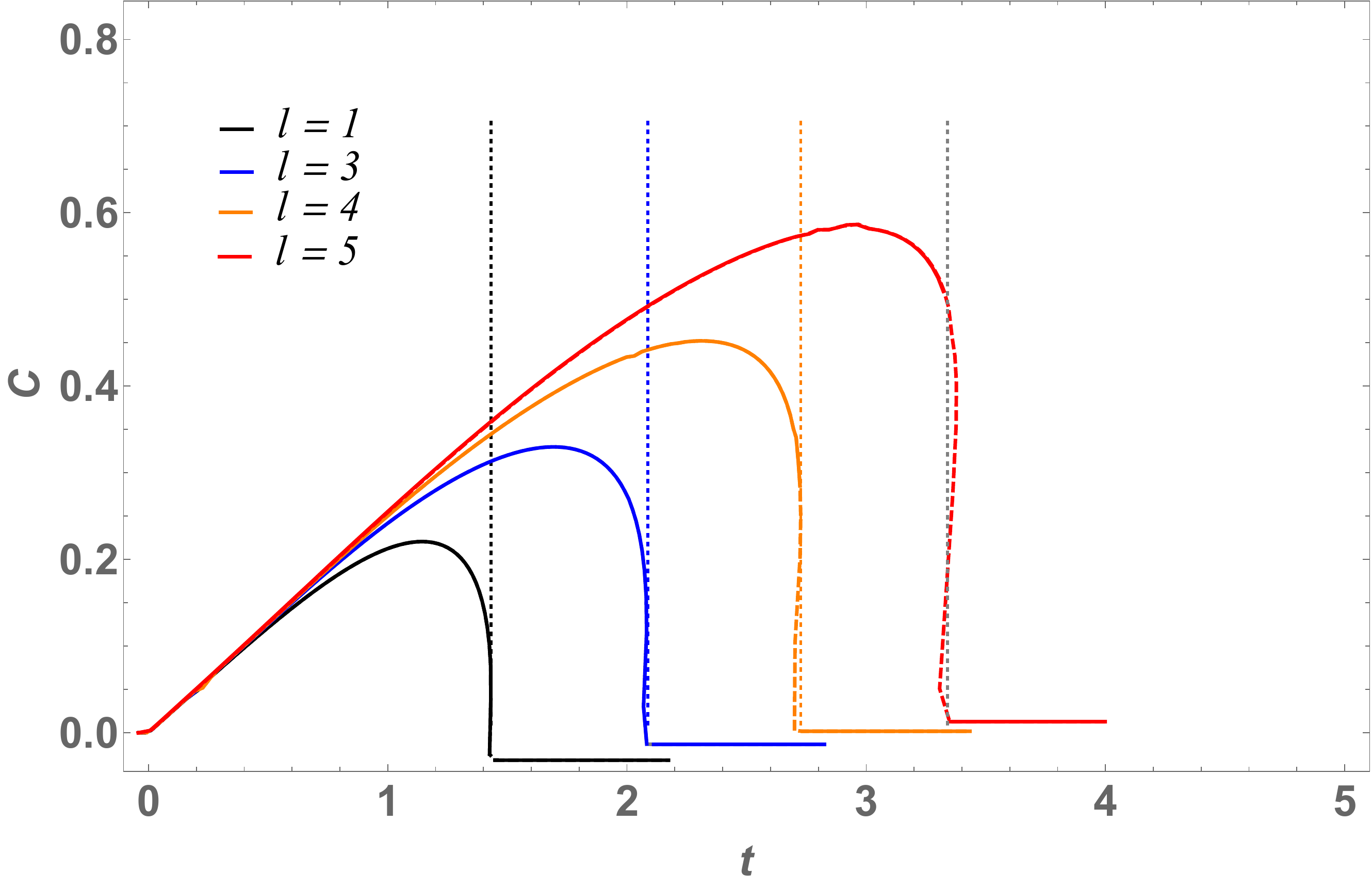}\ \hspace{1cm}
        \caption{The evolution of HEE and HC with samples of  stripe width for $\beta=1$.}
 \label{fig:q0b1}
\end{figure}

For non-vanishing $\beta$,  we find that  the effect of momentum relaxation for wider strip  is more explicit than that for narrow strip by comparing Fig.\ref{fig:q0l2zsc} and Fig.\ref{fig:q0l5zsc}. More interesting properties can be observed, which are summarized as follows:
\begin{itemize}
\item The turning point $z_{*}/z_h$, relative HEE and HC all tend to be constants as time evolves Specially, as $\beta$ increases, all the relative renormalized stable values become smaller. But it should be noticed that after saturation, $z_{*}/z_h\sim 1$, so one could expect $\gamma_{\mathcal{A}}$ wraps the horizon completely and the stable value of HEE is equal to the thermal entropy (thermal entropy density \eqref{adstemperature} times the volume)\cite{Ryu:2006bv}. In our model, the thermal entropy density increases as $\beta$ increases because larger $\beta$ corresponds to lower $z_h$ (i.e., larger horizon radius $r_h$), so we could compute the stable HEE using basic thermodynamics and the stable value of HEE should become larger if $\beta$ is large\footnote{Via subtracting the divergent term $2L/\epsilon$, we have computed the original HEE, \eqref{eq-S0}, for the static background \eqref{eq-metric} which is the stable state under the thermalization. We obtain that the value of HEE indeed increase as $\beta$ which is consistent with that for the thermal entropy density.}. However, this phenomena is not conflict with our numerical results because here we describe the relative renormalized HEE (see eq.\eqref{subtractHEE}).

\item A novel feature is that as $\beta$ increases to be a certain value, the swallow tail behavior in HEE and the multi-valued region in HC disappear as shown in Fig.\ref{fig:q0l5zsc}.  In other words, the evolution of $\gamma_{\mathcal{A}}$ recovers to be continue. It implies that comparing to the case in Einstein gravity,  in this model with axion field, the system with wide strip can  emerge discontinue evolution of $\gamma_{\mathcal{A}}$. We note that we do not observe the discontinue recovers as we further increase $\beta$ as we can.

\item Fig.\ref{fig:q0l2zsc} shows that  for $l=2$, the larger $\beta$  promotes the system to become equilibrium.  That is to say, larger $\beta$ need less time to reach the stable state in thermalization. While for $l=5$ in Fig.\ref{fig:q0l5zsc}, larger $\beta$ corresponds to longer time to become stable. Moreover, larger $\beta$ in this case is related with lower stable value of both HEE and HC, which is also different from that $\beta$ suppresses the stable HEE but  promotes HC for $l=2$. This property suggests that if we fix $q$ in advance, the time that HEE or HC needs to reach the stable values may not monotonically depend on $\beta$ for fixed $l$, and vice versa.

\item The systems with large $l$ need more  time to reach stable than those with small $l$.  This rule can also be explicitly extracted from Fig.\ref{fig:q0b1} where we present the evolution of HEE and HC with different $l$ for fixed $\beta=1$. This feature was also observed in Einstein case \cite{Chen:2018mcc} and Einstein-Born-Infeld case \cite{Ling:2018xpc} which can be directly explained via the picture of entanglement tsunami proposed in \cite{Liu:2013iza,Liu:2013qca}.

\end{itemize}

Note that since the numeric instability for large $\beta$ and $l$, we only exhibit the results for $\beta\leq 3$ in Fig.\ref{fig:q0l5zsc}.
But we expect that it is enough to show the universal characteristic for large $\beta$.
In future, we would also like to improve our algorithm to work out the results for large $\beta$
such that we can confirm the universal characteristic for large $\beta$.
It is also interesting to analytically study HEE and HC in the large $\beta$ limit,
which shall help to confirm the universal characteristic.

It is interesting and significant to understand the above results from analytical perspective. It was proposed
in \cite{Liu:2013iza,Liu:2013qca} that many properties of the time evolution of HEE after a thermal quench can be analytically calculated with the picture of entanglement tsunami, and later the authors of \cite{Ling:2019ien} applies the same strategy  to study the time evolution of HC. As they addressed that their calculations were based on that the metric function satisfies some properties, one of which is the asymptotical behavior $f(z)\sim 1-M z^d+\cdots$ as $z\to0$. However, in our study, the asymptotical behavior is $f(z)\sim 1-\frac{1}{2}\beta^2 z^{d-1}+M z^{d}+\cdots$, and the subleading term is not $z^d$-order but $z^{d-1}$-order with nonvanishing $\beta$. Thus, the calculation and conclusion of \cite{Liu:2013iza,Liu:2013qca,Ling:2019ien} could not generalize to our case. We hope to develop the analytical study in our model in the near future.

\subsection{Evolution of HEE and HC in charged case}
In this section, we shall turn on the charge in this model and study the joined effect of $\beta$ and $q$ on the evolution of HEE and HC.

\begin{figure}[ht!]
 \centering
  \includegraphics[width=7cm]{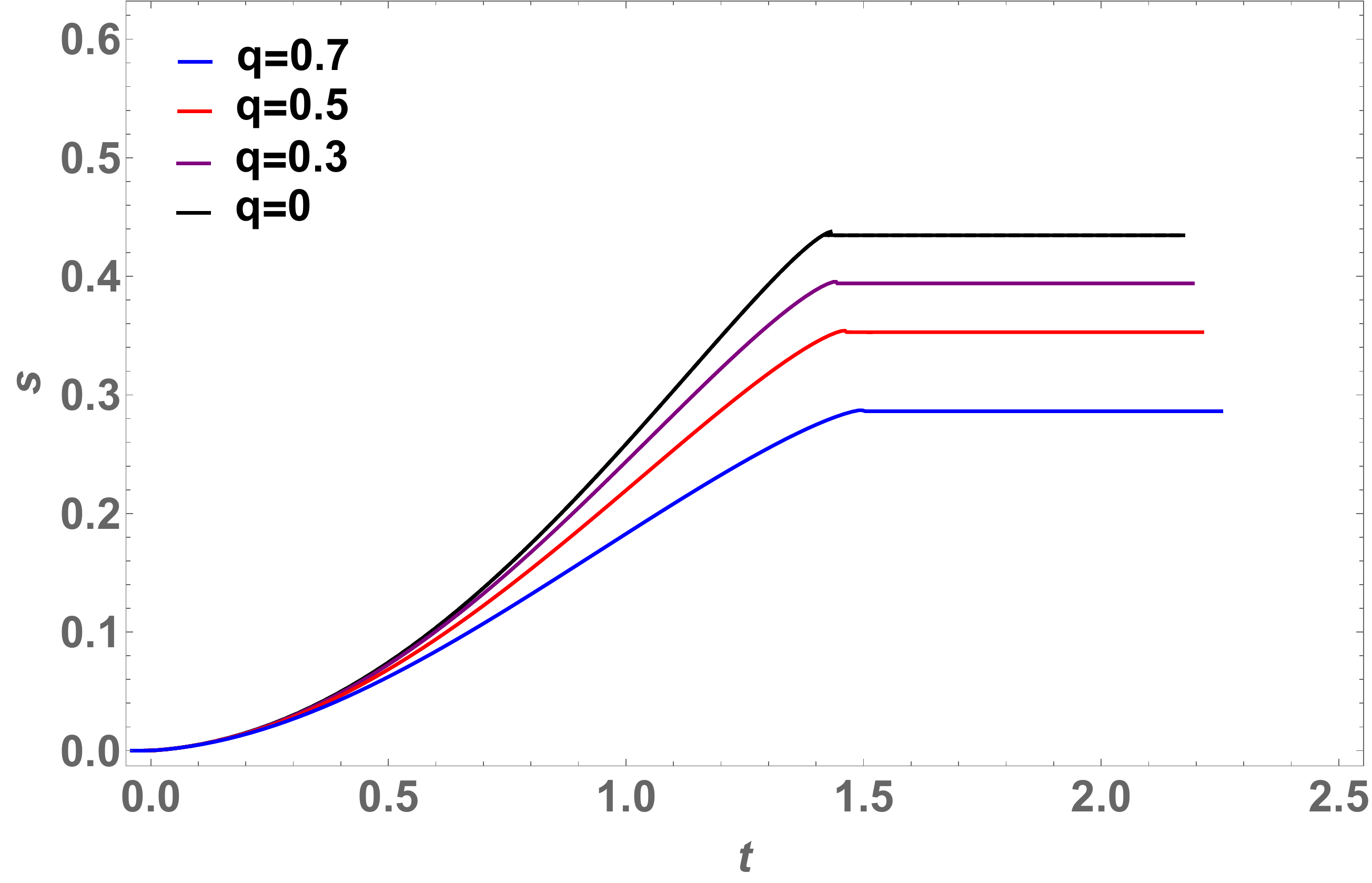}\ \hspace{0.1cm}
   \includegraphics[width=7cm]{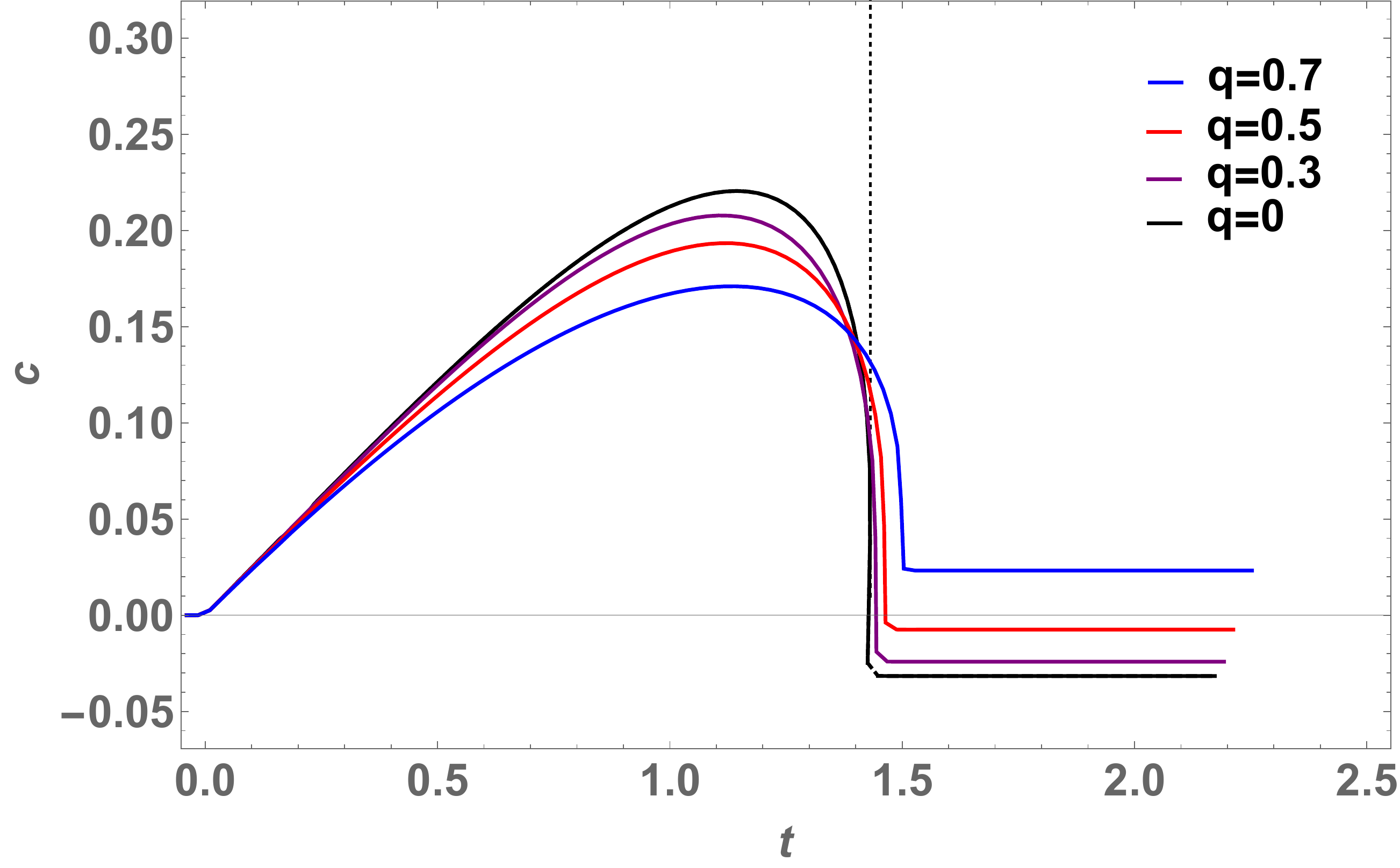}
	  \caption{The evolution of HEE (left panel) and HC (right panel) for different $q$ with $l$=2, $\beta =1$.}
 \label{fig:b1l2zsc}
\end{figure}

\begin{figure}[ht!]
 \centering
  \includegraphics[width=7cm]{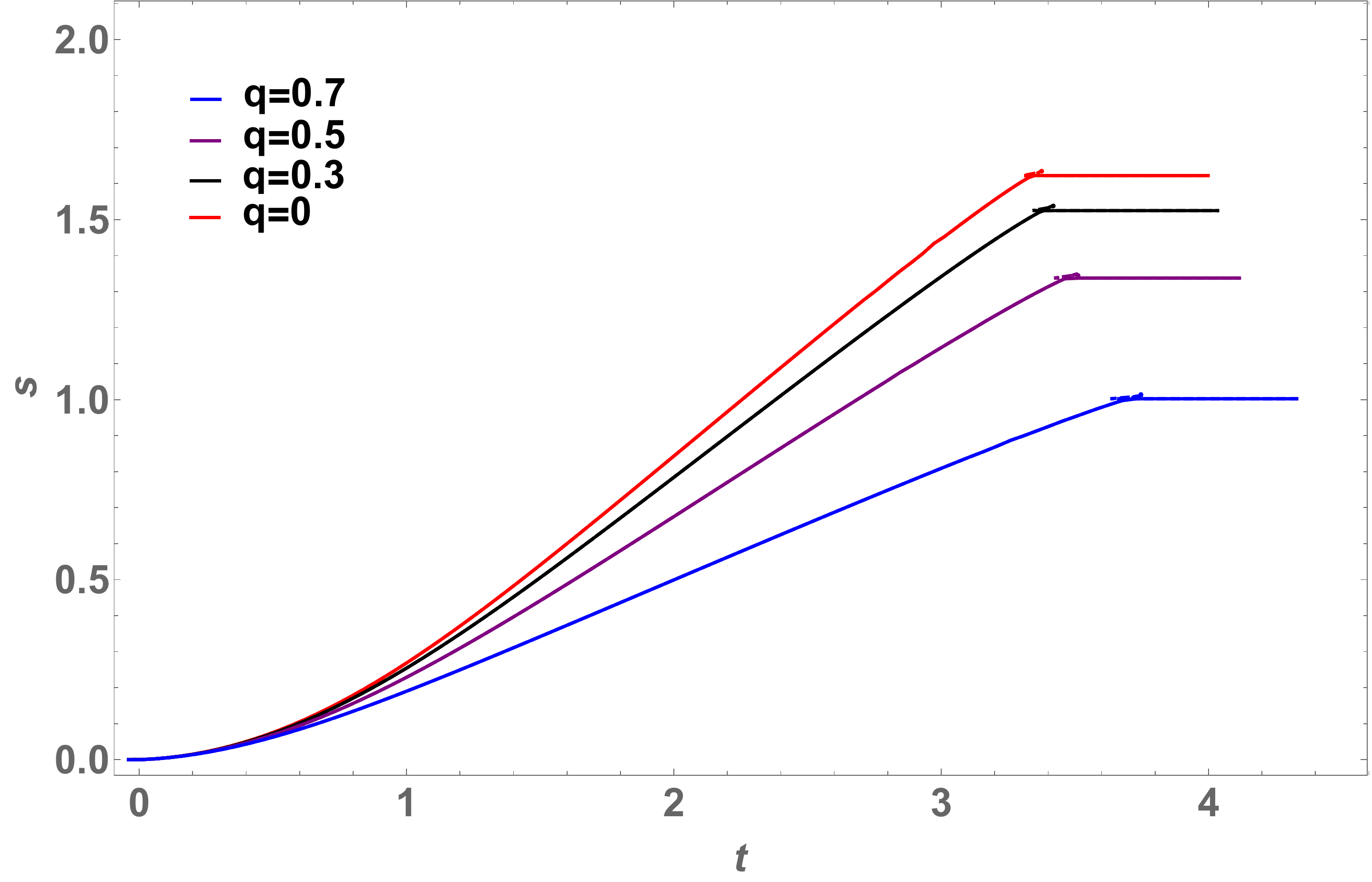}\ \hspace{0.1cm}
  \includegraphics[width=7cm]{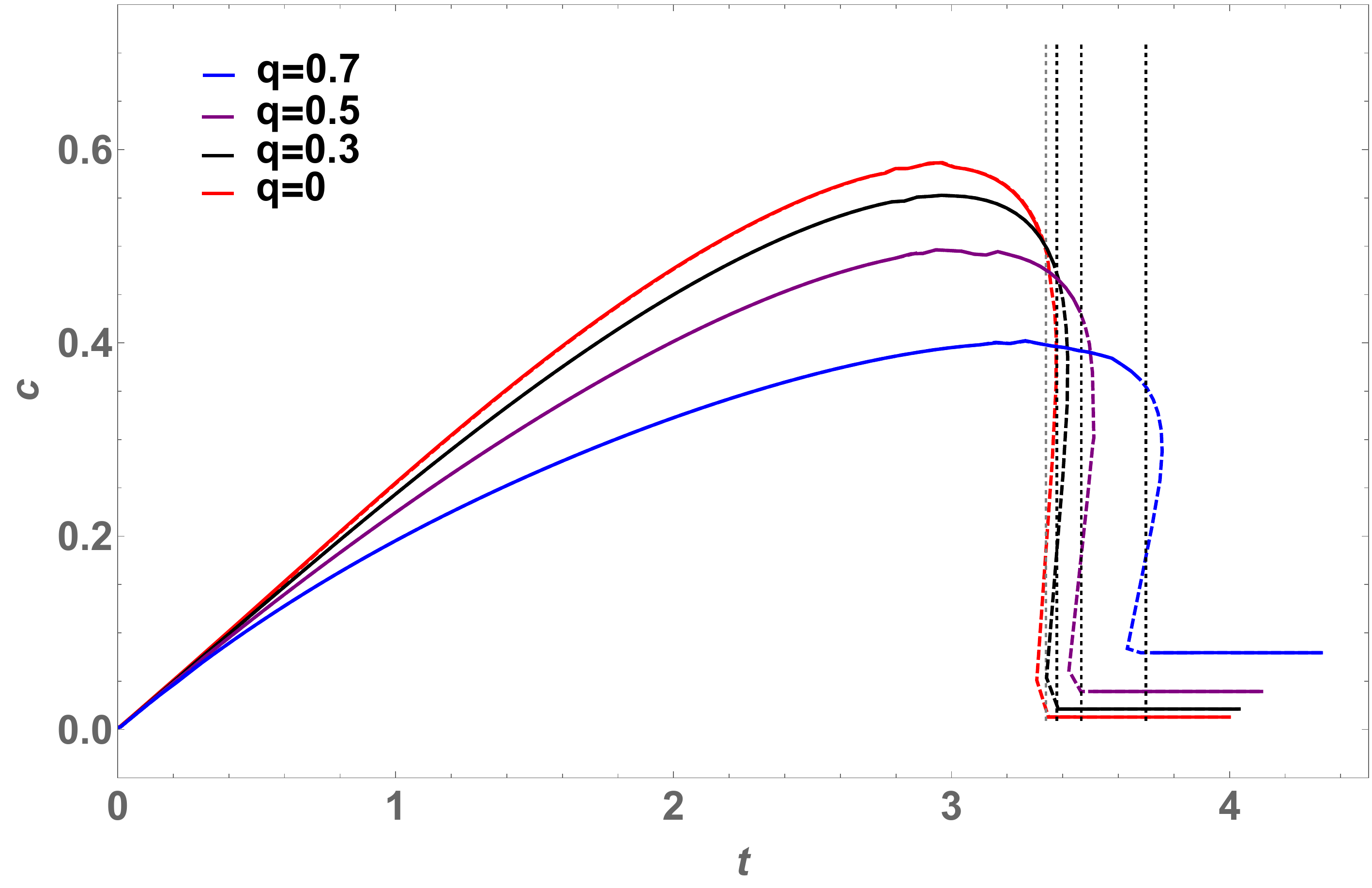}
	  \caption{The evolution of HEE (left panel) and HC (right panel) for different $q$ with $l$=5, $\beta =1$. The shallow tail behaviors and multi-valued regions are clearer than $l=2$ case.}
 \label{fig:b1l5zsc}
\end{figure}

It is noticed that the evolution of HEE and HC affected by the charge in four dimensional  Einstein theory and Einstein-Born-Infeld theory has been studied in \cite{Ling:2018xpc}. We first repeat their results and then turn on $\beta=1$ to see the effect of charge in the model with momentum relaxation. Our results are shown in Fig.\ref{fig:b1l2zsc} for a small width $(l=2)$  and  in Fig.\ref{fig:b1l5zsc} for a large width $(l=5)$. From the figures, we see that the charged black holes need longer time to become stable, and larger charge corresponds to smaller stable HEE but bigger stable HC. This phenomena are similar as the existed observes in \cite{Ling:2018xpc}.

It is worthwhile to point out  that for some parameters in our study, the final saturated HC could be almost zero and even negative, which seems to be universal behavior in the study of quench process with different models \cite{Chen:2018mcc,Ling:2018xpc,Zhang:2019vgl,Zhou:2019jlh,Ageev:2018nye}. However, this is in contrast to the behavior that the complexity is always
a monotonically increasing function of time found in \cite{Chapman:2018dem}. To understand this behavior could intrigue open questions: One is whether the normalized subregion complexity described via holography is dual to the complexity of formation of a state, which measures how difficult to map a given reference state into a desired state. The other is what is the reference state in the holographic study. Since the aforementioned questions are still open, so the negative equilibrium HC, which means that the stable value of the complexity is smaller than its initial value, could be acceptable but called for further understanding.

Comparing Fig.\ref{fig:b1l2zsc} and Fig.\ref{fig:b1l5zsc}, it is obvious that, again, larger size of width brings in swallow-tails in HEE and multi-values in HC due to the discontinue behavior of minimum area. Moreover, Fig.\ref{fig:b1l5zsc} shows that larger $q$ makes the discontinue behavior more evident, which is similar to that found in \cite{Ling:2018xpc,Zhou:2019jlh}.

\begin{figure}[ht!]
 \centering
  \includegraphics[width=7cm]{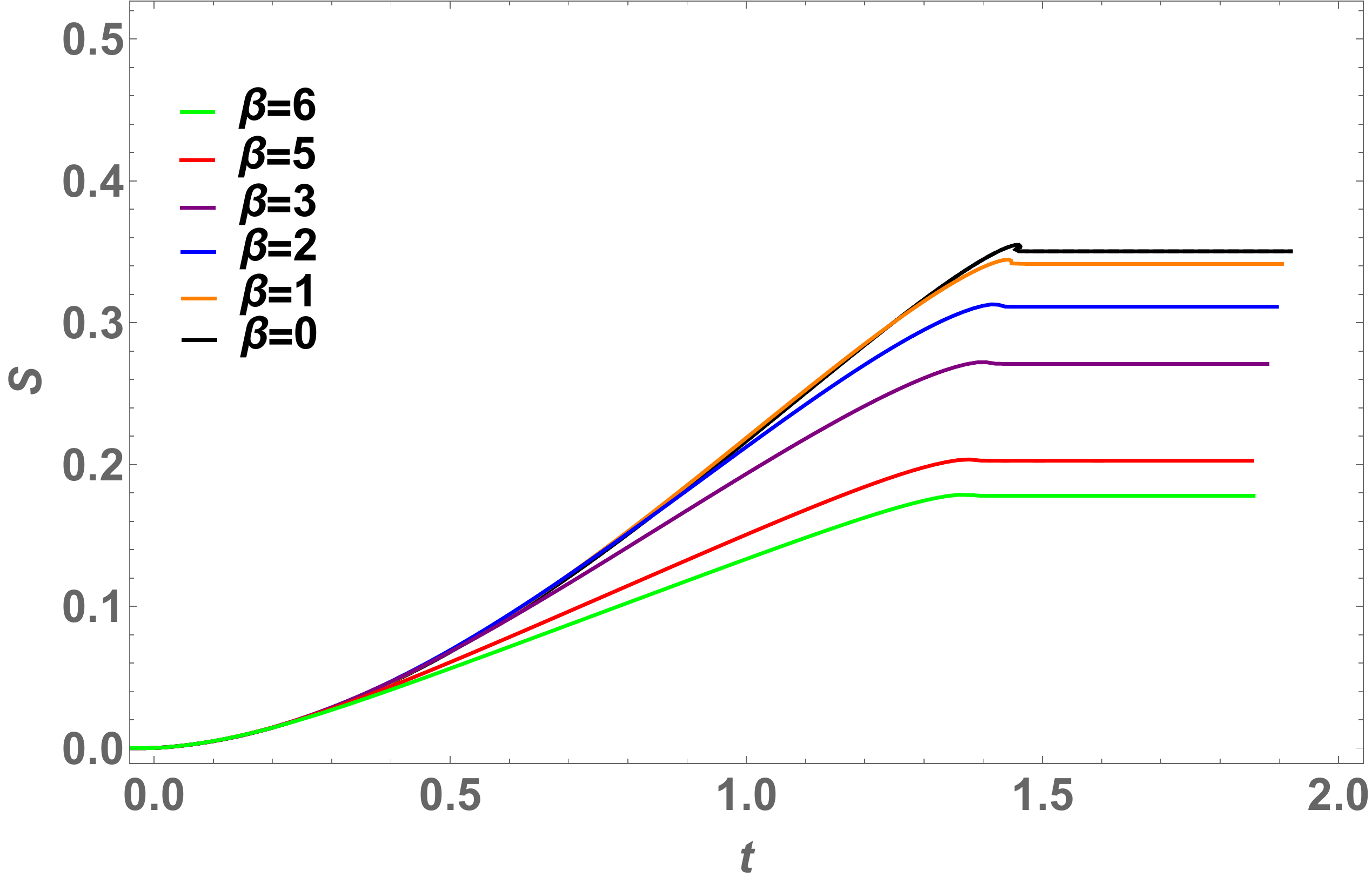}\ \hspace{0.1cm}
  \includegraphics[width=7cm]{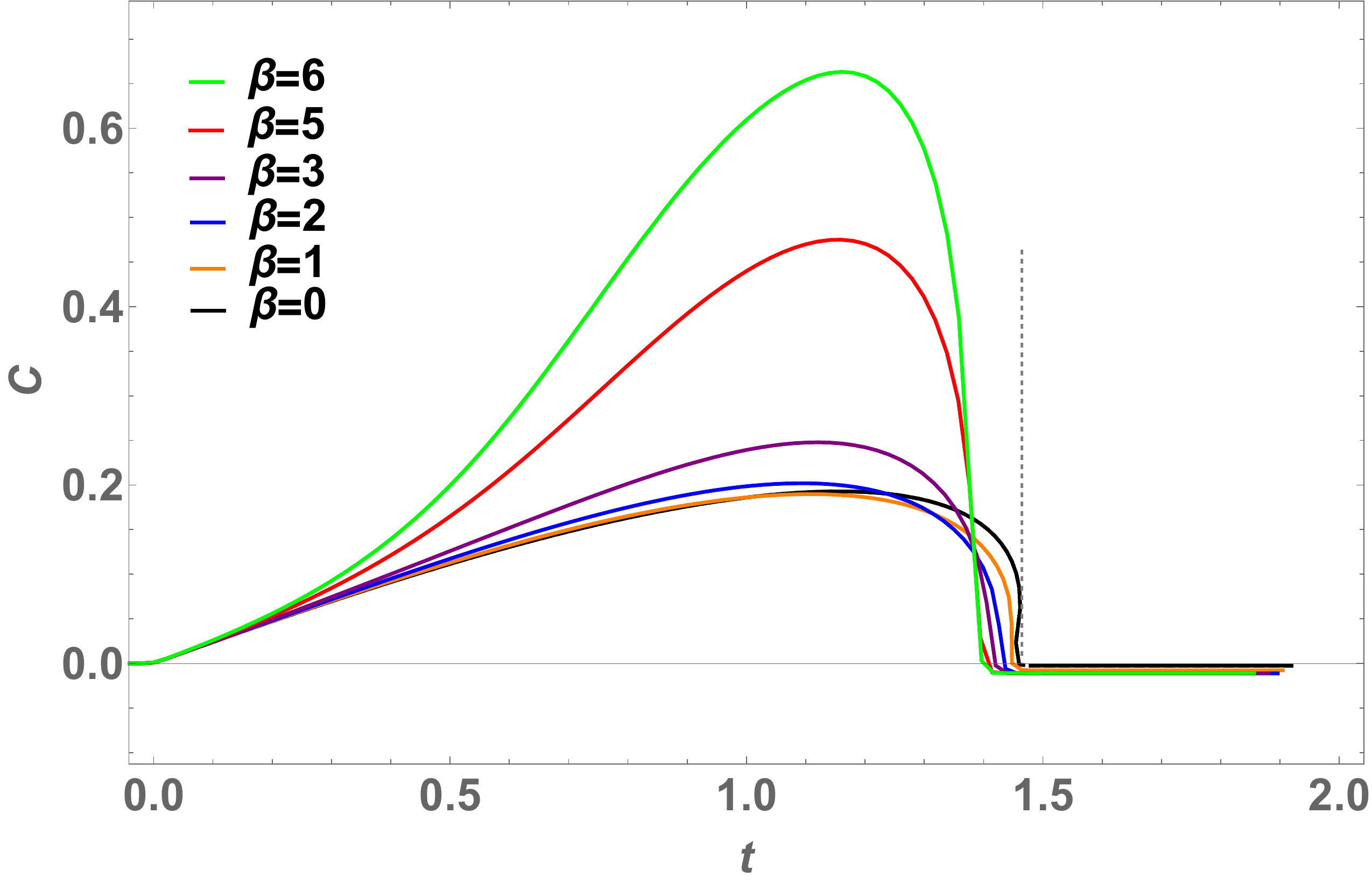}
	  \caption{Left: The evolution of HEE  the for different $\beta $ . Right: The evolution of HC for different $\beta $.
	  Here we have set $l$=2, $q=0.5$.}
 \label{fig:q05l2zsc}
\end{figure}

\begin{figure}[ht!]
 \centering
  \includegraphics[width=7cm]{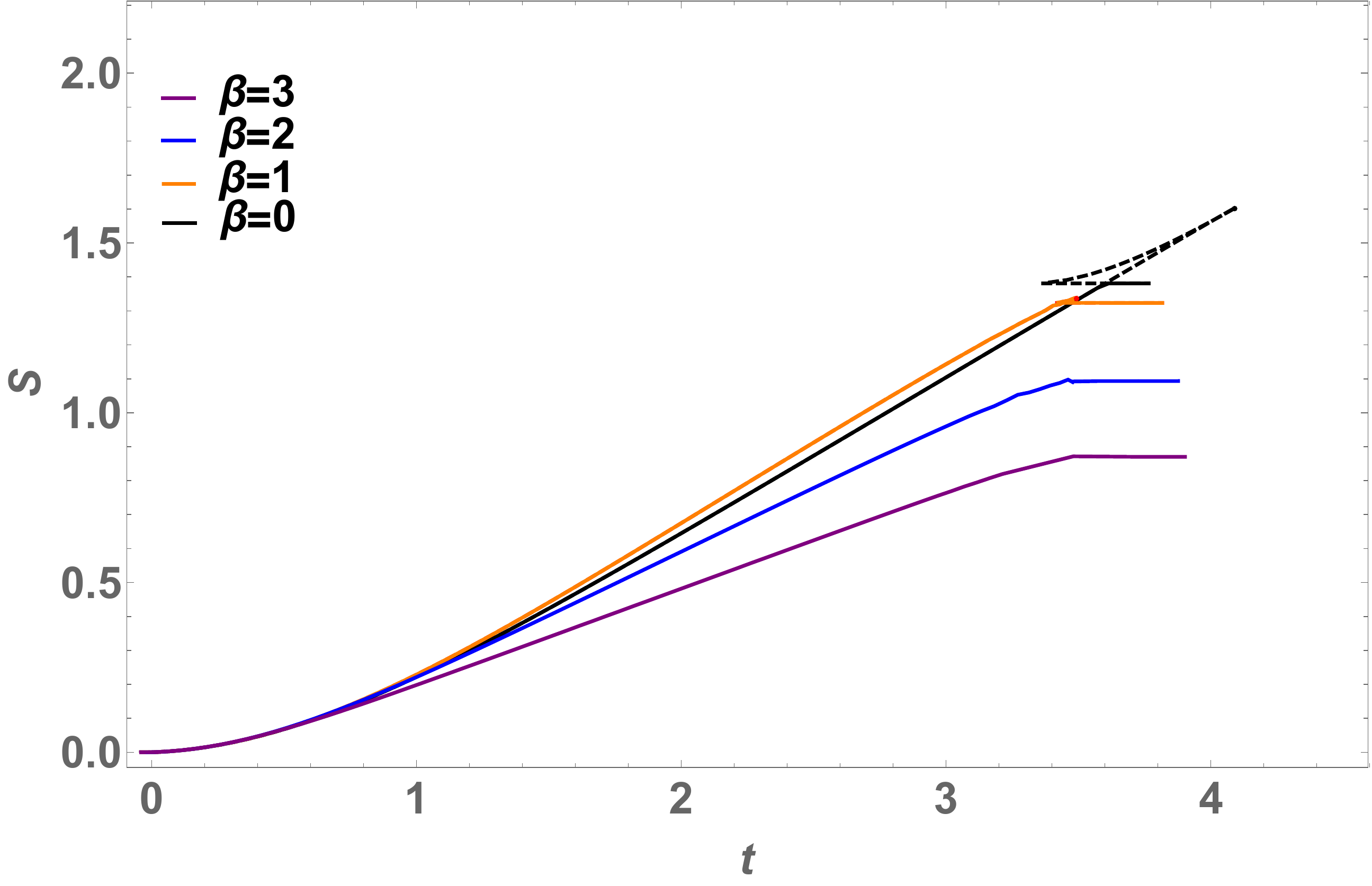}\ \hspace{0.1cm}
  \includegraphics[width=7cm]{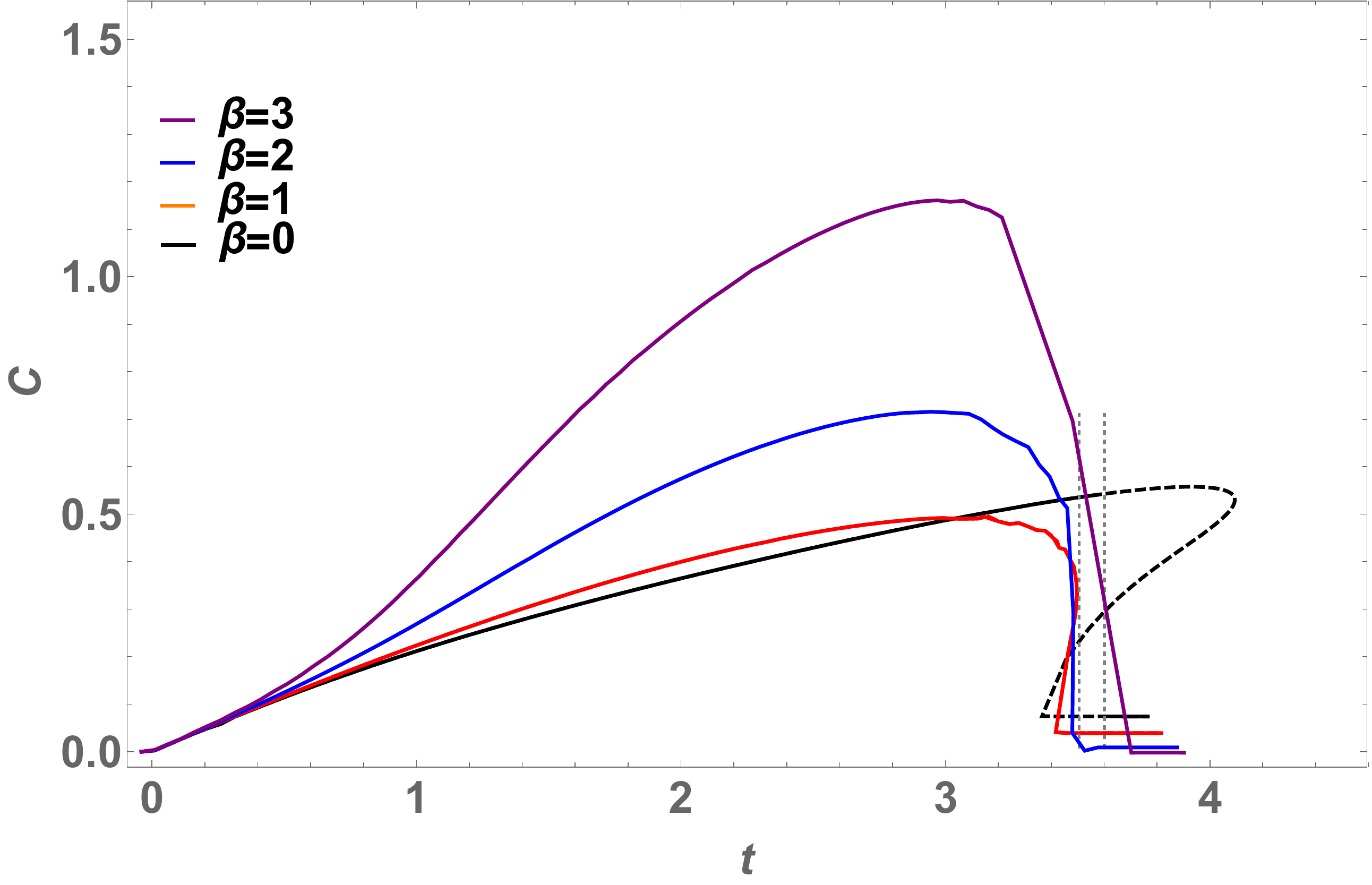}
	  \caption{Left:  The evolution of HEE for different $\beta $. Right: The evolution of HC for different $\beta $.
	  Here we have set $l$=5, $q=0.5$. Clearly, larger $\beta$ restrains the swallow tails and multi-valued regions.}
 \label{fig:q05l6zsc}
\end{figure}

In the charged case with $q=0.5$, we show the evolution of HEE and HC for different $\beta$ in Fig.\ref{fig:q05l2zsc} ($l=2$) and  Fig.\ref{fig:q05l6zsc} ($l=5$). For $l=2$, the evolution is still continue as shown in Fig.\ref{fig:q05l2zsc}. While in Fig.\ref{fig:q05l6zsc}, they are discontinue  and  $\beta$ also suppresses the effect of $l$, so that the swallow tail and multi-values behavior disappear as we increase $\beta$, which is similar to what we observe in neutral case.

\section{Conclusion and discussion}\label{sec:conclu}
In this paper, we studied the evolution of HEE and HC under a thermal quench in EMA theory. In this theory, the black brane solution is homogeneous but the dual  boundary  theory has momentum dispersion because the spacial dependent axion fields provide the source to break the translation symmetry.
We  mainly investigated  the effects of the momentum relaxation  and the charge of the black brane on the evolution of HEE and HC.

In neutral case,  when $l$ is tuned large, the evolutions of the HEE and the HC behave from a continuous function into a discontinuous one. For the continuous case,  the HEE increases at the first stage and then it arrives at a stable final region, while HC grows until it arrives at a maximum point, and then after that it quickly drops to become a stable final stage. For the discontinuous case, a swallow tail appears in the evolution of HEE while correspondingly, a multi-valued behavior can be seen in HC. This picture is the same as that found in \cite{Chen:2018mcc,Ling:2018xpc}.  However, as the momentum relaxation increases, the swallow tail behavior in HEE and the multi-valued region in HC disappear, i.e. the continuous  evolutions will be recovered. This denotes that in this model, the system with larger width would emerge discontinuous evolution of $\gamma_{\mathcal{A}}$ in contract to that in Einstein theory. Moreover, large $\beta$ corresponds to smaller stable value of both HEE and HC, while  the time that HEE or HC needs to reach the stable values is not monotonically depend on $\beta$ for fixed $l$, and vice versa. It would be interesting to understand our results from analytical approach.

We found that the charge would make the discontinue behaviour in HEE and HC more explicit at large $l$. That's to say, for bigger charge, the system with narrower size could emerge discontinuity in the evolution. This phenomena has also  been observed  in \cite{Ling:2018xpc,Zhou:2019jlh}. Similar as that occurred in neutral case, when we increase the momentum relaxation, the swallow tail in HEE and the multi-valued region in HC become weaken and finally the continuous  evolutions would be recovered.

In all of the evolutions we plotted, it is obvious that at the early stage the behaviour of evolution are almost the same, i.e., they  almost does not depend on the parameters including the size the the strip. This is because the growth of complexity stems from  the local operator excitations.   The possible bound of HEE growth rate for various stages of evolution has been discussed in \cite{Liu:2013qca}, while the Lloyd bound for HC growth rate $dC(t)/dt\leq 2M/\pi$, where $M$ is total mass of the system at any time $t$, has been addressed in \cite{Brown:2015bva}. Thus, it would be very interesting to carefully analyze the growth rate in various stages of the evolution of HEE and HC in our model, and compare them with the related bounds.
This  work is under progress.

Besides the HEE and HC,  our studies  can be extended into the related quantities we mentioned in the introduction, i.e., EoP, CoP and logarithmic negativity and  their evolution under the thermal quench using similar methods. EoP of mixed state in the dual theory of this model with momentum relaxation has been studied in \cite{Huang:2019zph} very recently. The evolutions of EoP in this setup of  Einstein gravity can be seen in \cite{Yang:2018gfq}, and it is natural to ask how its evolution will be affected when the momentum dispersion is involved in the system.

\begin{acknowledgments}
We appreciate Cheng-Yong Zhang for helpful discussions. 
This work is supported by the Natural Science Foundation 
of China under Grants No. 11705161, No. 11775036, and 
No. 11847313, and Fok Ying Tung Education Foundation 
under Grant No. 171006. X.-M. K. is also supported by the 
Natural Science Foundation of Jiangsu Province under
Grant No. BK20170481. J.-P.W. is also supported by Top 
Talent Support Program from Yangzhou University.
\end{acknowledgments}

\end{document}